\documentclass[11pt,a4paper]{article}
\usepackage[utf8]{inputenc}
\usepackage{braket}
\usepackage{amsmath}
\usepackage[english,russian]{babel}
\usepackage{graphicx}
\graphicspath{ {imageseps/} }
\usepackage{hyperref}
\hypersetup{
    colorlinks=true,
    linkcolor=blue,
    filecolor=magenta,
    urlcolor=cyan,
}
\usepackage{tikz-feynman}
\usepackage[nottoc,numbib]{tocbibind}

\begin{document}

\selectlanguage{english}


\vspace{5mm}

\centerline{\Large \bf Secularly growing loop corrections to the dynamical Casimir effect}

\vspace{5mm}

\centerline{S. O. Alexeev}

\begin{center}
{\it Institutskii per, 9, Moscow Institute of Physics and Technology, 141700, Dolgoprudny, Russia}
\end{center}

\begin{center}
{\it B. Cheremushkinskaya, 25, Institute for Theoretical and Experimental Physics, 117218, Moscow, Russia}
\end{center}

\vspace{3mm}



\centerline{\bf Abstract}
The paper is based on the Bachelor Thesis defended this year in ITEP, Moscow. This is the extended version of \cite{Akhmedov:2017hbj} [arXiv:1707.02242] and contains a lot more technical details of the calculations. We consider (1+1)-dimensional massless scalar field theory with Dirichlet boundary conditions on arbitrary time-like curve. It is well known that in this situation there is a non-zero energy flux at the tree-level, if the latter curve corresponds to a non--stationary motion of the boundary. Such a problem is usually referred to as the radiation due to moving mirrors. We calculate quantum loop corrections to the energy flux from moving mirrors and find that they grow with time. Hence, they are not suppressed in comparison with the semi--classical contributions. Thus, we observe the break down of the perturbation theory, discuss its physical origin and ways to deal with such a situation.

\vspace{4mm}

\renewcommand{\contentsname}{\centering Contents}
\tableofcontents

\newpage

\section{Introduction}\

It appears that quantum field theory in non--stationary situations and in the presence of a medium is quite poorly understood in comparison with that in the vacuum state.  Namely, it frequently happens that the physical phenomena that appear in non--stationary situations are quite counterintuitive, if one accepts the common wisdom which is gained during the study of the vacuum quantum field theory. E.g. it is a commonly accepted opinion that the flux which is generated by a moving mirror is saturated by its semi--classical value \cite{BirDav}, \cite{FullDav}. The goal of the present paper is to show that this is not the case in a self--interacting theory.

At the same time, it is known in condensed matter theory that in non--stationary situations in non--Gaussian theories semi--classical approximation breaks down (see e.g. \cite{LL} and \cite{Kamenev}). The same situation is observed in de Sitter space quantum field theory \cite{AkhmedovKEQ}--\cite{Akhmedov:2013vka}), in the scalar QED on the strong electric field backgrounds \cite{Akhmedov:2014hfa}, \cite{Akhmedov:2014doa} and in the quantum corrections to the Hawking radiation \cite{AkhGodPop}.

Non-stationarity leads to the absence of time homogeneity, which, in turn, leads to the violation of the energy conservation law. That happens in non--closed system. As a result, there are such processes, which are forbidden otherwise. For instance, in interacting theories non-stationarity causes loop corrections to propagators to grow with time, which means that even if the coupling constant is very small, the loop corrections to physical quantities become comparable to the tree-level contributions during long enough time evolution. To understand the physics of secularly growing corrections fully, it is necessary to make a resummation of at least leading contributions coming from all loops. For the system considered in this paper the question of resummation is still an open issue. In comparison with other systems with secularly growing loop corrections, this one appears to be more complex: as it is shown in the main text of the paper, in addition to the fact that in this case two-loop corrections to the two-point correlation functions grow with time non-linearly, loop corrections to four-point correlation functions also demonstrate the secular growth, and all these effects make it unclear how to produce the resummation.

The main text consists of 10 parts. In the section \ref{Modes} we formulate the problem and find mode functions, solving the wave equation with null boundary condition on some time-like curve. In the section \ref{CanonicalCommutationRelation} we check if the canonical commutation relation between the field operator $\phi$ and its conjugated momentum $\pi$ is satisfied. In the section \ref{EMT} we rederive the well-known formula for the energy flux in the tree-level approximation. In the section \ref{Wightman} the Wightman function at coincident points is computed. The section \ref{FreeHamiltonian} contains the discussion of the free Hamiltonian in the presence of mirrors. The sections \ref{GeneralExpForLoops}-\ref{OneLoop} contain the details of the calculations of the loop corrections to the Keldysh propagator. Finally, the section \ref{FourPoint} contains the calculation of the loop corrections to the four-point correlation functions.

\section{Mode functions} \label{Modes}
\quad\quad We consider the massless, two--dimensional, real scalar field theory:

$$
S = \int d^2x \, \left[\partial_\mu \phi \, \partial^\mu \phi - \frac{\lambda}{4} \, \phi^4\right],
$$
with the field $\phi(t,x)$ being equal to $0$ on some time-like curve $\big(t, z(t)\big)$, which is a world-line of some massive object, called a mirror:
$$\phi\big(t,z(t)\big)=0.$$
We assume that the boundary terms in the action are trivial.\\

Let us first study the tree-level approximation $\lambda=0$. For any mirror world-line $z(t)$ the field operator $\phi(t,x)$ can be represented in the following form:
$$\phi(t,x)=\int_{0}^{\infty} \frac{dk}{2\pi} \big[a_\omega h_k(t,x)+\operatorname{h.c.}\big],$$
where $a_k,\ a_{k'}^\dag$ are annihilation and creation operators, satisfying the standard commutation algebra: $[a_k,a_{k'}^\dag]=2\pi\delta(k-k').$
Mode functions $h_k(t,x)$ solve Klein-Gordon equation and satisfy the same boundary condition as the field $\phi$: $$h_k\big(t, z(t)\big)=0.$$
In this section we find modes $h_k(t,x)$ for arbitrary mirror world-line $z(t)$.

\subsection{Modes for the mirror at rest}
\begin{figure}[t]
\centering
\includegraphics[scale=0.25]{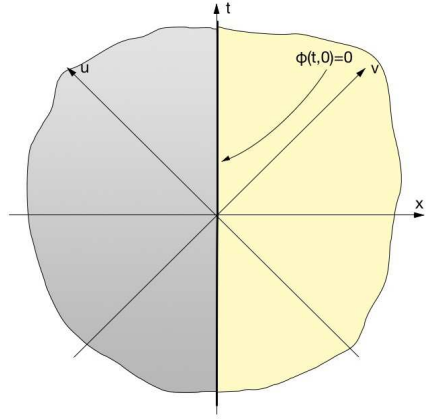}
\caption{The world line of a mirror at rest. The field $\phi$ is equal to $0$ on it. QFT is considered in the region to the right hand side of the mirror.}
\label{fig:1}
\end{figure}

\quad\quad Let us start with the case when the mirror is always at rest, i.e. with the world-line $z(t)\equiv0$. To quantize the theory we solve the free wave equation ($\lambda = 0$) with the following boundary conditions (see figure \ref{fig:1}):

\begin{equation}
\label{eq:1}
\partial_\mu \partial^\mu \phi \equiv \partial^2 \phi (t,x)=0,\quad {\rm and} \quad \phi(t,0)=0.
\end{equation}
Any solution to this equation can be represented in the following form:
$$\phi(t,x)=\int_{0}^{\infty} \frac{dk}{2\pi} \sin(k x) \big[A_k e^{-i k t}+A_k^* e^{ikt}\big],$$
where $A_k$ is a function of $k$. So, the quantized field $\phi(t,x)$ takes the form:
$$\hat \phi(t,x)=\int_{0}^{\infty} \frac{dk}{2\pi} \sin(k x) \big[A_k e^{-i k t}\  \hat a_k+\operatorname{h.c.}\big],$$
where the creation and annihilation operators $a_k, a_k^\dag$ satisfy the commutation relations: \begin{equation}\label{eq:commuta}[a_{k'},a_k^\dag]=2\pi\delta(k-k').\end{equation}
Requiring the field operator and its conjugate momentum to satisfy the canonical commutation relation, we obtain the coefficient $A_k$. On the one hand,
$$[\phi(x,t), \ \partial_{t}\phi(y,t)]=i\delta(x-y).$$
On the other hand,
$$[\phi(x,t),\partial_t \phi(y,t)]=2\int_{0}^{\infty} \frac{dk}{2\pi}\  (i k) |A_k|^2 \sin(k x) \sin(k y).$$
Defining $A_k=\sqrt{2/k}$, we obtain:
\begin{equation}
\label{eq:2}
\begin{split}
[\phi(x,t),\partial_t \phi(y,t)]=&-i \int_{0}^{\infty} \frac{dk}{2\pi} \big[ e^{ik(x+y)}-e^{ik(x-y)}-e^{-ik(x-y)}+e^{-ik(x+y)} \big]=\\
&=-i \int_{-\infty}^{\infty} \frac{dk}{2\pi} \big[ e^{ik(x+y)}-e^{ik(x-y)}\big]=i\delta(x-y)-i\delta(x+y).
\end{split}
\end{equation}
In addition to the expected term $i\delta(x-y)$, we obtain the term $i\delta(x+y)$, which is not equal to $0$ only when $x=y=0$, i.e. when both points are on the boundary, and, hence, we can drop it off.\\

Thus, modes $h_k$ have the following form:
$$h_k(t,x)=\sqrt{\frac{2}{k}}\operatorname{sin}(kx)e^{-ikt}.$$
In the light-cone coordinates $v=t+x,\ u=t-x$:
\begin{equation}
\label{eq:3}
h_k(u,v)=\frac{i}{\sqrt{2k}}\big[e^{-ikv}-e^{-iku}\big].
\end{equation}
The first exponential function $e^{-ikv}$ just represents a plane wave, moving towards the mirror, while the second one $e^{-iku}$ is a wave, reflected from the mirror in such a way to provide the null boundary conditions on it.

\subsection{Modes for the mirror moving with constant velocity}
\quad\quad To find the modes in this case
\begin{equation}
\label{eq:4}
\partial^2 \phi (t,x)=0,\quad {\rm and} \quad \phi(t,-\beta t)=0,
\end{equation}
one should just find a reflected wave, which provides the right boundary conditions. It is obvious that the following modes
\begin{equation}
\label{eq:5}
h_k(u,v)=\frac{i}{\sqrt{2k}}\big[e^{-ikv}-e^{-ik\alpha u}\big]
\end{equation}
satisfy the wave equation and are equal to $0$ on the line $(t, -\beta t)$, with $\alpha$ being equal to $\displaystyle \frac{1-\beta}{1+\beta}$.

\begin{figure}[t]
\centering
\includegraphics[scale=0.25]{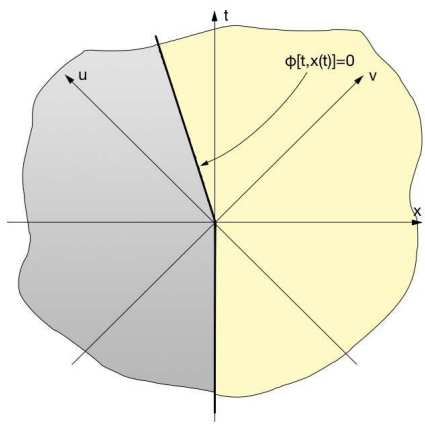}
\caption{A broken world-line. Reflected waves in regions $u>0$ and $u<0$ are different.}
\label{fig:3}
\end{figure}

\subsection{Modes for a broken world-line}
\quad\quad Let us continue with the case, when the mirror stays at rest until $t=0$ and then instantaneously starts its motion with a constant speed (see figure \ref{fig:3}):

\begin{equation}
\label{eq:6}
z(t)=\begin{cases}
\ \ 0 & t<0\\

  -\beta t& t\ge0.
\end{cases}
\end{equation}
In such a situation we just have different reflected waves in different space-time regions:

\begin{equation}
\label{eq:7}
h_k(u,v)=\frac{i}{\sqrt{2k}}\big[e^{-ikv}-\theta(-u)e^{-ik u}-\theta(u)e^{-ik \alpha u}\big].
\end{equation}
It is straightforward to check that these modes obey the Klein-Gordon equation and satisfy the necessary boundary conditions.

\subsection{Modes for an arbitrary world-line} \

\begin{figure}[h]
\centering
\includegraphics[scale=0.25]{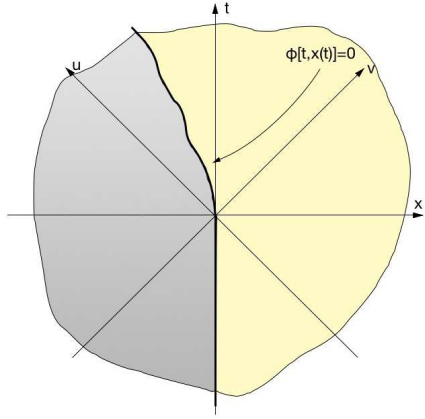}
\caption{An arbitrary smooth world-line.}
\label{fig:4}
\end{figure}

If we know the solution to the wave equation with null boundary conditions on the broken world-line (figure \ref{fig:3}), then it is possible to obtain from it a solution, which is equal to $0$ on an arbitrary smooth world-line (figure \ref{fig:4}): We are going to approximate the world-line by a broken curve, write modes for this discretization and then take the limit when the broken curve becomes smooth.\\

So, let us consider the following world-line: $z(t)\equiv 0,\ t<0$ and when $t\ge 0$ this world-line consists of straight lines connecting points $(t_i, z_i)$ and $(t_{i-1},z_{i-1}),\ i=1, \dots,\infty$, with $(t_0, z_0)$ being equal to $(0,0)$. We also suppose that $t_i-t_{i-1}=\Delta t$ for all values of $i$ and that $z_i<z_{i-1}$. We know the solution to KG equation in this case to be as follows:
$$e^{-ikv}-\theta(-u)e^{-iku}-\sum_{i=1}^{\infty}\theta\big[-(t-x)+t_i-z_i\big]\cdot \theta\big[t-x-(t_{i-1}-z_{i-1})\big]\cdot \operatorname{exp}\bigg[-i\omega\bigg(\alpha_i u +\frac{z_{i-1} t_i-z_i t_{i-1}}{\Delta t}(1+\alpha_i)\bigg)\bigg],$$
with $\alpha_i$ standing for
$$\frac{1-\beta_i}{1+\beta_i},\  \operatorname{where}\ \beta_i=-\frac{z_i-z_{i-1}}{\Delta t}\equiv-\frac{\Delta z_i}{\Delta t}\ge 0.$$
The series written above consists of exponents, which solve KG equation and provide the correct boundary conditions in different regions of space-time, given by the theta-functions. Now we expand all functions of $t_{i-1}$ near $t_i$ to the first order in $\Delta t$:

\begin{equation*}
\begin{split}
&\sum_{i=1}^{\infty}\theta\big[-(t-x)+t_i-z_i\big]\cdot \theta\big[t-x-(t_{i-1}-z_{i-1})\big]\cdot \operatorname{exp}{\bigg\{-i\omega\bigg(\alpha_i u +(1+\alpha_i)\frac{z_{i-1} t_i-z_i t_{i-1}}{\Delta t}\bigg)\bigg\}}=\\
&=\sum_{i=1}^{\infty}\theta\big[-(t-x)+t_i-z_i\big]\cdot \bigg( \theta\big[t-x-(t_{i}-z_{i})\big]+\delta\big[t-x-(t_i-z_i)\big]\bigg(1-\frac{\Delta z_i}{\Delta t}\bigg) \Delta t\bigg)\times \\
&\times \operatorname{exp}\bigg\{-i\omega\bigg[\alpha_i u +(1+\alpha_i)\bigg(z_i-\frac{\Delta z_i}{\Delta t} t_i\bigg)\bigg]\bigg\}+O(\Delta t^2).
\end{split}
\end{equation*}

Using the fact that $\theta(x)\theta(-x)=0$, we simplify this sum to:
$$\theta(0)\sum_{i=1}^{\infty} \Delta t\ \delta\big[t-x-(t_i-z_i)\big]\bigg(1-\frac{\Delta z_i}{\Delta t}\bigg)\cdot \operatorname{exp}\bigg\{-i\omega\bigg[\alpha_i u +(1+\alpha_i)\bigg(z_i-\frac{\Delta z_i}{\Delta t} t_i\bigg)\bigg]\bigg\}+O(\Delta t^2).$$
Taking the limit $\Delta t \to 0$, we almost obtain the final result:
$$\theta(0)\int_0^{\infty}  d\tau\ \delta\big[u-(\tau-z(\tau)\big]\bigg(1-\frac{dz}{d\tau}\bigg)\cdot \operatorname{exp}\bigg\{-i\omega\bigg[\alpha(\tau) u +\bigg(1+\alpha(\tau)\bigg)\bigg(z(\tau)-\frac{dz}{d\tau} \tau\bigg)\bigg]\bigg\}=\theta(0)\cdot e^{-i\omega (2t_u-u)},$$
where $t_u$ is implicitly defined by the equation: $u=t_u-z(t_u)$. Comparing this result to what one should obtain in the case $z(t)\equiv 0,\  t>0$ it follows that we should put $\theta(0)=1$ here.\\

Thus, for an arbitrary mirror world-line $z(t)$ we have the following modes (see \cite{BirDav},\cite{FullDav}):

\begin{equation}
\label{eq:8}
h_k(u,v)=\frac{i}{\sqrt{2k}}\big[e^{-ikv}-e^{-ik(2t_u-u)}\big],
\end{equation}
where $t_u$ is a solution of the equation: $t_u-z(t_u)=u$.

\section{Canonical commutation relations} \label{CanonicalCommutationRelation}\

In this section we check that above defined modes (\ref{eq:8}) provide the correct commutation relation between the field operator $\phi$ and the conjugate momentum $\pi=\partial_t \phi$:

\begin{equation*}
\begin{split}
& \phi(u,v)=\int_{0}^{\infty} \frac{dk}{2\pi} \big[a_k h_k(u,v)+\operatorname{h.c.}\big],\\
& \partial_t \phi=\int_{0}^{\infty}\frac{dk}{2\pi}\ \sqrt{\frac{k}{2}} \bigg[a_k\bigg[e^{-ik v}-e^{-ik (2t_u-u)}\bigg(2\frac{dt_u}{du}-1\bigg)\bigg]+\operatorname{h.c.}\bigg].
\end{split}
\end{equation*}
Substituting these expressions into the commutor $[\phi, \pi]$ and using eq. (\ref{eq:commuta}), we obtain:
$$[\phi (t,x),\ \partial_t \phi(t,y)]=\frac{i}{2}\int_{0}^{\infty}\frac{dk}{2\pi}\bigg[ \big[e^{-ik v_x}-e^{-ik(2t(u_x)-u_x)}\big]\big[e^{ik v_y}-e^{ik(2t(u_y)-u_y)}f(u_y)\big]+\operatorname{h.c.}\bigg],$$
where $$f(u)=2\frac{dt_u}{du}-1=\frac{2}{1-z'(t_u)}-1=\frac{2}{1+\beta(t_u)}-1=\frac{1-\beta(t_u)}{1+\beta(t_u)}$$
(we suppose that $z'(t)=-\beta(t)$ and $\beta(t)\geq0$). Therefore,
\begin{equation*}
\begin{split}
& [\phi (t,x),\ \partial_t \phi(t,y)]=\frac{i}{2}\bigg\{\delta\big[v_x-v_y\big]+f(u_y)\cdot\delta\big[2t(u_x)-u_x-2t(u_y)+u_y\big]-\\
& -f(u_y)\cdot\delta\big[v_x-2t(u_y)+u_y\big]-\delta\big[2t(u_x)-u_x-v_y\big]\bigg\}=\\
& =\frac{i}{2}\bigg[\delta(v_x-v_y)+\frac{f(u_y)}{|f(u_y)|}\delta(u_x-u_y)+\operatorname{boundary}\ \operatorname{terms}\bigg]=i\delta(x-y)+\operatorname{boundary}\  \operatorname{terms}.
\end{split}
\end{equation*}
"Boundary terms"\ stand for the contributions, which are proportional to $f(u_y)\delta\big[v_x-2t(u_y)+u_y\big]+\delta\big[2t(u_x)-u_x-v_y\big]$. The latter $\delta$-functions are not equal to $0$ only when $x=y=z(t)$, i.e. when both points simultaneously lie on the boundary. So, in the region to the right of the boundary, we obtain the standard commutation relation:
\begin{equation}
\label{eq:commut}
[\phi(t,x),\pi(t,y)]=i\delta(x-y).
\end{equation}

\section{The expectation value of the energy-momentum tensor (tree-level contribution)} \label{EMT}\

In this section we study how the mirror motion influences the energy flux at tree-level (described by the ${tx}-$component of the energy-momentum tensor). For the beginning, we reproduce the well known formula for a general case of a mirror motion \cite{BirDav}. Then we obtain the answers for some concrete cases.

\subsection{A general world-line} \

To calculate $T_{tx}$, we use the following procedure to regularize divergent integrals \cite{FullDav}:

\begin{equation}
\label{eq:ttxcommon}
\braket{T_{tx}}=\lim_{\varepsilon\to 0}\  \frac{1}{2}\left\langle \partial_t\phi(t,x)\partial_x\phi(t+i\varepsilon,x)^{\phantom{\frac12}} + \,\, \partial_x\phi(t,x)\partial_t\phi(t+i\varepsilon,x)\right\rangle,\end{equation}
where the average is taken with respect to the ground state $\ket{0}$, corresponding to the modes under consideration (such that $a_k\ket{0}=0$ for all $k$). Substituting the expression of the field operator $\phi(t,x)$ through the modes $h_k(u,v)$ into this formula, we obtain the following expression:
$$\braket{T_{tx}}=\lim_{\varepsilon\to 0}\  \frac{1}{4\pi}\int_{0}^{\infty} dk\ k\big[e^{-k\epsilon}-p'(u)p'(u+i\varepsilon)e^{ik[p(u+i\varepsilon)-p(u)]}\big],$$
where $p(u)=2\tau_u-u$ and $\displaystyle p'(u)=\frac{dp}{du}$. Performing integration and taking the limit as $\epsilon$ goes to $0$, we obtain the formula, valid for an arbitrary mirror world-line. Furthermore, the result can be expressed in terms of mirror velocity \cite{FullDav}:
\begin{equation}\braket{T_{tx}} (u)=\frac{1}{24\pi}\bigg[\frac{p^{'''}}{p'}-\frac{3}{2}\bigg(\frac{p^{''}}{p^{'}}\bigg)^2\bigg]=-\frac{1}{12\pi}\  \frac{(1+v)^{1/2}}{(1-v)^{3/2}}\  \frac{d}{dt}\frac{\dot{v}}{(1-v^2)^{3/2}}\  \bigg|_{t=\tau_u}.\end{equation}
One can notice that $\displaystyle \frac{\dot{v}}{(1-v^2)^{3/2}}$ is the mirror acceleration in its instantaneous rest frame.

\subsection{Examples} \

Let us look for the applications of the formula for $T_{tx}$, that was obtained in the previous subsection.\\

If the mirror moves with a constant speed, then its proper acceleration is equal to $0$, so the formula gives $T_{tx}=0$.\\

In the case of the broken world-line the equation under consideration does not work, because the derivative of $p(u)$ is not defined. Nevertheless, we can easily obtain $\braket{T_{tx}}$ from straightforward calculations. In this case we should take modes from eq. (\ref{eq:7}). Their spatial and time derivatives are:
$$\partial_t h_k(u,v)=(\partial_v+\partial_u)h_k(u,v)=\sqrt{\frac{k}{2}}\bigg[\theta(u)\big(e^{-ik v}-\alpha e^{-ik \alpha u}\big)+\theta(-u)\big(e^{-ik v}-e^{-ik u}\big)\bigg],$$
$$\partial_x h_k(u,v)=(\partial_v-\partial_u)h_k(u,v)=\sqrt{\frac{k}{2}}\bigg[\theta(u)\big(e^{-ik v}+\alpha e^{-ik \alpha u}\big)+\theta(-u)\big(e^{-ik v}+e^{-ik u}\big)\bigg].$$
Now we substitute these derivatives into the eq. (\ref{eq:ttxcommon}) for $T_{tx}$ :
\begin{equation}
\begin{split}
& \braket{T_{tx}} = \frac{1}{4\pi} \lim_{\varepsilon\to 0}\bigg\{\int_{0}^{\infty} k dk\ \bigg[  \theta(u)\theta(u+i\varepsilon)\big(e^{-k\varepsilon}-\alpha^2e^{-k\alpha\varepsilon}\big) + \theta(-u)\theta(-u-i\varepsilon)\big(e^{-k\varepsilon}-e^{-k\varepsilon}\big)\bigg] + \nonumber \\
& + \int_{0}^{\infty} k dk\ \bigg[  \theta(u)\theta(-u-i\varepsilon)\big(e^{-k\varepsilon}-\alpha e^{-ik\alpha u} e^{ik u}e^{-k\varepsilon}\big)+\theta(-u)\theta(u+i\varepsilon)\big(e^{-k\varepsilon}-\alpha e^{-ik u} e^{ik\alpha u}e^{-k\alpha\varepsilon}\big)\bigg]\bigg\} = \nonumber \\
& = \frac{1}{4\pi} \lim_{\varepsilon\to 0}\bigg\{ \theta(-u)\theta(u+i\varepsilon)\bigg[\frac{1}{\varepsilon^2} +
\frac{\alpha}{\big[u(\alpha-1)+i\alpha\varepsilon\big]^2}\bigg] + \theta(u)\theta(-u-i\varepsilon)\bigg[\frac{1}{\varepsilon^2} + \frac{\alpha}{\big[u(1-\alpha)+i\varepsilon\big]^2}\bigg] \bigg\}= \nonumber \\
& = \frac{1}{4\pi} \lim_{\varepsilon\to 0}\bigg\{\theta(-u)\bigg[\theta(u)+\delta(u)i\varepsilon - \frac{1}{2}\delta'(u)\varepsilon^2+O(\varepsilon^3)\bigg]\bigg[\frac{1}{\varepsilon^2} + \frac{\alpha}{u^2(\alpha-1)^2}+O(\varepsilon)\bigg] + \nonumber \\
& +\theta(u)\bigg[\theta(-u) - \delta(u)i\varepsilon + \frac{1}{2}\delta'(u)\varepsilon^2 + O(\varepsilon^3)\bigg]\bigg[\frac{1}{\varepsilon^2} + \frac{\alpha}{u^2(\alpha-1)^2} + O(\varepsilon)\bigg]\bigg\} = \nonumber \\
& = \frac{1}{4\pi} \lim_{\varepsilon\to 0}\bigg\{\frac{1}{2}\delta'(u)\big[\theta(u) - \theta(-u)\big] + O(\varepsilon)\bigg\}=\frac{1}{8\pi}\delta'(u)\big[\theta(u) - \theta(-u)\big].
\end{split}
\end{equation}
Despite the fact that this result is defined only as a generalized function, it meets the physical requirements, according to which it is not equal to $0$ only when $u=0$.\\

\begin{figure}[h]
\centering
\includegraphics[scale=0.25]{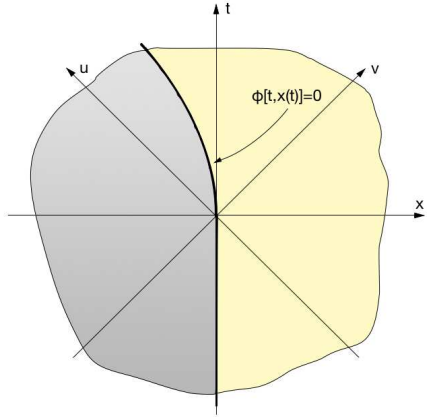}
\caption{The world-line of the mirror, approaching the speed of light.}
\label{fig:5}
\end{figure}

We are also interested to obtain $\braket{T_{tx}}$ for the world-line, which approaches the light-like line as $t$ goes to infinity (see figure \ref{fig:5}):

\begin{equation}
\label{eq:exptraj}
z(t)=\begin{cases}
\ \ 0 & t<0\\

  -t+a(1-e^{-t/a})& t\ge0,
\end{cases}\end{equation}
where $1/a$ is proportional to the mirror proper acceleration. To obtain $\braket{T_{tx}}$ in this case we use the general formula. Let us first calculate the derivatives of $p(u)$:
$$p(u)\equiv2t_u-u\equiv t_u+z(t_u)=a\big(1-e^{-t_u/a}\big),$$
$$p'(u)\equiv \frac{dp}{du}=\frac{dp}{dt_u}\frac{dt_u}{du}={\bigg(1-\frac{dz}{dt_u}\bigg)}^{-1}\frac{dp}{dt_u}=\frac{e^{-t_u/a}}{2-e^{-t_u/a}}=\frac{1}{2e^{t_u/a}-1},$$
$$p''(u)=-\frac{2}{a}\cdot\frac{e^{-t_u/a}}{(2-e^{-t_u/a})^3}, \ p'''(u)=\bigg(\frac{2}{a}\bigg)^2\frac{e^{-t_u/a}(1+e^{-t_u/a})}{(2-e^{-t_u/a})^5}.$$
As the result,
$$\braket{T_{tx}}=\frac{1}{12\pi a^2}\frac{2e^{-t_u/a}-1}{(e^{-t_u/a}-2)^4}.$$
When $u$ goes to infinity, the energy flux becomes almost constant and equal to $\displaystyle -\frac{1}{2^4 \cdot12\pi a^2}.$

\section{The Wightman function (tree-level contribution)} \label{Wightman}
\quad\quad In this section we consider the tree-level behaviour of the Wightman function at the coincident points for different mirror world--lines:
\begin{equation*}
\begin{split}
& D^W(x,y)\big|_{x\to y} \equiv \braket{\phi(t,x)\phi(t+i\epsilon,x)}= \int_{0}^{\infty} \frac{dk}{2\pi}\frac{1}{2k}\left[e^{-ikv}-e^{-ikp_u}\right]\left[e^{ik(v+i\epsilon)}-e^{ikp(u+i\epsilon)}\right]=\\
& =\int_{0}^{\infty} \frac{dk}{2\pi}\frac{1}{2k}\left[e^{-k\epsilon}+e^{-k\epsilon p'(u)}-e^{-k\big(\epsilon+i(p_u-v)\big)}-e^{-k\big(\epsilon p'(u)+i(v-p_u)\big)}\right]=
\end{split}
\end{equation*}
\begin{equation*}
\begin{split}
& =\frac{1}{4\pi}\log{\frac{\left[\epsilon+i(p_u-v)\right]\left[\epsilon p'(u)+i(v-p_u)\right]}{\epsilon^2 p'(u)}}=\frac{1}{4\pi}\log\bigg[1+\frac{i(v-p_u)}{\epsilon p'(u)}+\frac{i(p_u-v)}{\epsilon}+\frac{(p_u-v)^2}{\epsilon^2 p'(u)}\bigg]\approx\\
& \approx \frac{1}{4\pi} \log\frac{(p_u-v)^2}{\epsilon^2p'(u)},\quad {\rm as} \quad \epsilon \to 0.
\end{split}\end{equation*}

First, consider the case of the mirror's world--line, approaching asymptotically the speed of light, (\ref{eq:exptraj}). For this world-line $p(u)$ and its derivative in the limit $u\to \infty$ are as follows:
\begin{equation*}\begin{split}& p(u)=a\big(1-e^{-t_u/a}\big)\approx a,\\
& p'(u)=\frac{1}{2e^{t_u/a}-1}\approx \frac{e^{-u/2a}}{2}.\end{split}\end{equation*}
Thus, in this case the Wightman propagator at coincident points grows linearly in $u$ and logarithmically in $v$:
\begin{equation}D^W(x,y)\big|_{x\to y}\approx \frac{1}{4\pi}\log \frac{2e^{u/2a}(a-v)^2}{\epsilon^2}= \frac{1}{4\pi}\bigg[\frac{u}{2a}+\log\frac{2(a-v)^2}{\epsilon^2} \bigg],\ u\to \infty.\end{equation}

Second, consider the following world-line ($\beta<1$):
\begin{equation*}
z(t)=\begin{cases}
\ \ 0 & t<0\\

  -\beta t+a(1-e^{-\beta t/a})& t\ge0,
\end{cases}\end{equation*}
which describes the mirror which approaches, as $t \to + \infty$, a velocity which is less than speed of light, $\beta < 1$. Then, $p(u)$ and $p'(u)$ for $u\to \infty$ are as follows:
\begin{equation*}\begin{split}& p(u)=t_u(1-\beta)+a\big(1-e^{-\beta t_u/a}\big)\approx t_u(1-\beta)\approx \alpha u,\\
& p'(u)=\frac{1-\beta(1- e^{-\beta t_u/a})}{1+\beta(1- e^{-\beta t_u/a})}\approx \frac{1-\beta}{1+\beta}\equiv\alpha.\end{split} \end{equation*}
So, for this world-line the Keldysh propagator at coinciding points grows logarithmically as the function of $(\alpha u -v)$:
\begin{equation}D^W(x,y)\big|_{x\to y}\approx \frac{1}{4\pi}\log \frac{\big(\alpha u-v\big)^2}{\alpha\epsilon^2},\ u\to \infty.\end{equation}
Such a behavior of the Wightman function is very similar to the one of the massless minimally coupled scalar field in the four--dimensional de Sitter space--time: see e.g. \cite{Starobinsky} and \cite{Tsamis:2005hd} for the explanations and extensions. We will see below that the loop contributions in the case under study have also very similar behaviour to those in the case of the massless scalars in de Sitter space.

\section{The free Hamiltonian}  \label{FreeHamiltonian}\

In this section we find the free Hamiltonian for various mirror world-lines and show that for non-stationary mirror world-lines it can not be diagonalized once and forever. Nevertheless, some linear combination of the Hamiltonian $H$ and the momentum operator $P$, which is an evolution operator along the direction of the mirror world line, can be diagonalized when the motion of the mirror becomes stationary.

\subsection{The mirror at rest} \

Let us calculate the free Hamiltonian, integrating the energy density over the spatial coordinate:

\begin{equation}
\label{eq:hamiltcommon}
H^0=\frac{1}{2}\int_{0}^{\infty} dx \big[(\partial_t\phi)^2+(\partial_x\phi)^2\big].\end{equation}
Substituting the field operator, expressed in terms of the modes from eq. (\ref{eq:3}), into this formula, we obtain:
$$H^0=\frac{1}{2}\int_{0}^{\infty}\frac{dk}{2\pi}\int_{0}^{\infty}\frac{dk'}{2\pi}\sqrt{kk'}\int_{0}^{\infty}dx\bigg[\left(a_k a_{k'}+a_k^{\dag}a_{k'}^{\dag}\right)\left(e^{-i(k+k')v}+e^{-i(k+k')u}\right)+$$
$$+\left(a_k a_{k'}^{\dag}+a_k^{\dag}a_{k'}\right)\left(e^{-i(k-k')v}+e^{-i(k-k')u}\right)\bigg].$$
The first term gives $\delta(k+k')$ after the integration over $x$, the second one $-$ $\delta(k-k')$. Consequently,
\begin{equation}
\label{eq:hamilt1}H^0=\frac{1}{2}\int_{0}^{\infty}\frac{dk}{2\pi}\ k\left[a_k a_{k}^{\dag}+a_k^{\dag}a_{k}\right].\end{equation}
We see that the modes we have found diagonalize the free Hamiltonian.

\subsection{The mirror moving with a constant velocity} \

Using eq. (\ref{eq:hamiltcommon}) and the modes ({\ref{eq:5}}), we find that in this case the Hamiltonian depends on time:
$$H=\frac{1}{4}\int_{0}^{\infty}\frac{d\omega}{2\pi}(1+\alpha)\omega\bigg[a_\omega a_\omega^{\dag}+a_\omega^{\dag}a_\omega\bigg]+$$
\begin{equation}\label{eq:hamiltonianconst}+\frac{i}{2}(\alpha-1)\int_{0}^{\infty}\frac{d\omega}{2\pi}\int_{0}^{\infty}\frac{d\omega'}{2\pi}\frac{\sqrt{\omega\omega'}(\omega+\omega')}{(\omega+\omega')^2+0^2}\bigg[a_\omega a_{\omega'} e^{-it(1-\beta)(\omega+\omega')}-a_\omega^{\dag} a_{\omega'}^{\dag} e^{it(1-\beta)(\omega+\omega')}\bigg]+\end{equation}
$$+\frac{i}{2}(\alpha-1)\int_{0}^{\infty}\frac{d\omega}{2\pi}\int_{0}^{\infty}\frac{d\omega'}{2\pi}\frac{\sqrt{\omega\omega'}(\omega-\omega')}{(\omega-\omega')^2+0^2}\bigg[ a_\omega a_{\omega'}^{\dag} e^{-it(1-\beta)(\omega-\omega')}-a_\omega^{\dag} a_{\omega'}e^{it(1-\beta)(\omega-\omega')}\bigg].$$
It means that it can not be diagonalized once and forever. Also, we see that the operator of momentum has the same structure:
$$P=\frac{1}{2}\int_{-\beta t}^{\infty} dx\ \big[\partial_t\phi\ \partial_x\phi+\partial_x\phi\ \partial_t\phi\big]=\ \frac{1}{4}\int_{0}^{\infty}\frac{d\omega}{2\pi}(1-\alpha)\omega\bigg[a_\omega a_\omega^{\dag}+a_\omega^{\dag}a_\omega\bigg]+$$
\begin{equation}\label{eq:momentumconst}-\frac{i}{2}(\alpha+1)\int_{0}^{\infty}\frac{d\omega}{2\pi}\int_{0}^{\infty}\frac{d\omega'}{2\pi}\frac{\sqrt{\omega\omega'}(\omega+\omega')}{(\omega+\omega')^2+0^2}\bigg[a_\omega a_{\omega'} e^{-it(1-\beta)(\omega+\omega')}-a_\omega^{\dag} a_{\omega'}^{\dag} e^{it(1-\beta)(\omega+\omega')}\bigg]+\end{equation}
$$-\frac{i}{2}(\alpha+1)\int_{0}^{\infty}\frac{d\omega}{2\pi}\int_{0}^{\infty}\frac{d\omega'}{2\pi}\frac{\sqrt{\omega\omega'}(\omega-\omega')}{(\omega-\omega')^2+0^2}\bigg[ a_\omega a_{\omega'}^{\dag} e^{-it(1-\beta)(\omega-\omega')}-a_\omega^{\dag} a_{\omega'}e^{it(1-\beta)(\omega-\omega')}\bigg].$$
Since \begin{equation}
\label{eq:alpharel}
\alpha-1=-\frac{2\beta}{1+\beta}\  \operatorname{and}\ \alpha+1=\frac{2}{1+\beta},\end{equation}
we notice that the following linear combination of $H$ and $P$ has a very simple form and is diagonal:
\begin{equation}
\label{eq:lincomb}
H-\beta P=\frac{1}{2}\int_{0}^{\infty}\frac{d\omega}{2\pi}(1-\beta)\omega\bigg[a_\omega a_\omega^{\dag}+a_\omega^{\dag}a_\omega\bigg].\end{equation}
This result is not surprising, because the operator $H-\beta P$ defines translations along the direction of motion of the mirror (we mean space-time direction). In fact,
$$\exp\big(-iP_{\mu}x^{\mu}\big)\bigg|_{x=-\beta t}=\exp(-iHt-iPx)\bigg|_{x=-\beta t}=\exp\big[-it(H-\beta P)\big].$$
At the same time, as can be seen from the equations (\ref{eq:hamiltonianconst}) and (\ref{eq:momentumconst}), if $\beta \to 1$ (mirror approaches the speed of light) $H$ and $P$ are time independent and can be diagonalized separately.

\subsection{The broken world-line} \

In this case it is possible to represent modes in the following form (see the equation (\ref{eq:7})):
\begin{equation}
\label{eq:harmthet}
h_k(u,v)= \theta(u)h_k^\beta(u,v)+\theta(-u)h_k^0(u,v), \end{equation}\
where $$h_k^0(u,v)=\frac{i}{\sqrt{2k}}\big[e^{-ikv}-e^{-iku}\big],\quad {\rm and} \quad  h_k^\beta(u,v)=\frac{i}{\sqrt{2k}}\big[e^{-ikv}-e^{-ik\alpha u}\big].$$
We can also represent derivatives of the modes as follows:
$$\partial_t h_\omega(u,v)=\theta(u)\partial_t h_\omega^{\beta}+\theta(-u)\partial_t h_\omega^{0},$$
$$\partial_x h_\omega(u,v)=\theta(u)\partial_x h_\omega^{\beta}+\theta(-u)\partial_x h_\omega^{0}.$$
Similarly, the field operator $\phi$ and its derivatives can be rewritten in the same form:
$$\phi=\theta(u)\phi^{\beta}+\theta(-u)\phi^{0},$$
$$\partial_t \phi=\theta(u)\partial_t \phi^{\beta}+\theta(-u)\partial_t \phi^{0},$$
$$\partial_x \phi=\theta(u)\partial_x \phi^{\beta}+\theta(-u)\partial_x \phi^{0},$$
where $\phi^\beta$ contains expansion in $h^\beta$ modes, while $\phi^0$ --- in $h^0$ modes. From the relations above it follows that the Hamiltonian and the momentum operator have the following structure:
$$H=\theta(-t)H^{0}+\theta(t)\bigg[\int_{-\beta t}^t dx\ \big[(\partial_t \phi^{\beta})^2+(\partial_x\phi^\beta)^2\big]+\int_{t}^{\infty}dx\ \big[(\partial_t \phi^{0})^2+(\partial_x\phi^0)^2\big]\bigg],$$
$$P=\theta(-t)P^{0}+\theta(t)\bigg[\int_{-\beta t}^t dx\ \big[\partial_t\phi^{\beta}\partial_x\phi^{\beta}+\partial_x\phi^{\beta}\partial_t\phi^{\beta}\big]+\int_{t}^{\infty}dx\ \big[\partial_t\phi^{0}\partial_x\phi^{0}+\partial_x\phi^{0}\partial_t\phi^{0}\big]\bigg],$$
where $H_0$ and $P_0$ are the same Hamiltonian and momentum operator as in the case, in which the mirror is at rest.\\

Now we substitute expressions of the field $\phi$ and its derivatives through the operators $a_\omega,\ a_\omega^{\dag}$ into the expressions above:
$$H=\theta(-t)H^{0}+\theta(t)\iint \frac{d\omega}{2\pi}\frac{d\omega'}{2\pi}\frac{\sqrt{\omega\omega'}}{2}\bigg[a_\omega a_{\omega'}\bigg(e^{-i(\omega+\omega')t}\frac{e^{-i(\omega+\omega')x}}{-i(\omega+{\omega'})}\bigg|_{-\beta t}^t+\alpha^2e^{-i(\omega+\omega')\alpha t}\frac{e^{i(\omega+\omega')\alpha x}}{i(\omega+\omega')\alpha}\bigg|_{-\beta t}^t\bigg)+$$
$$+\ ...\bigg]+\theta(t)\iint \frac{d\omega}{2\pi}\frac{d\omega'}{2\pi} \frac{\sqrt{\omega\omega'}}{2} \bigg[a_\omega a_{\omega'}\bigg(\frac{e^{-2i(\omega+\omega')t}}{i(\omega+\omega')}-\frac{1}{i(\omega+\omega')}\bigg)+...\bigg].$$
The ellipsis denote similar terms with $a_\omega a_{\omega'}^{\dag},\ a_{\omega}^{\dag} a_{\omega'}$ and $a_{\omega}^\dag a_{\omega'}^\dag$.\\

The explicit form of the Hamiltonian is as follows:
$$H=\theta(-t)H^{0}+\theta(t)\iint \frac{d\omega}{2\pi}\frac{d\omega'}{2\pi}\frac{\sqrt{\omega\omega'}}{2}\big[a_\omega a_{\omega'}\  \xi(\omega, \omega')+a_\omega a_{\omega'}^{\dag}\ \xi(\omega, -\omega')+a_\omega^{\dag} a_{\omega'}\  \xi(-\omega,\omega')+a_\omega^\dag a_{\omega'}^\dag\ \xi(-\omega,-\omega')\big],$$
where
\begin{equation}\begin{split} & \xi(\omega,\omega')= e^{-i(\omega+\omega'-i\varepsilon)t}\frac{e^{-i(\omega+\omega'-i\varepsilon)x}}{-i(\omega+\omega'-i\varepsilon)}\bigg|_{-\beta t}^t+\alpha^2e^{-i(\omega+\omega'+i\varepsilon)\alpha t}\frac{e^{i(\omega+\omega'+i\varepsilon)\alpha x}}{i(\omega+\omega'+i\varepsilon)\alpha}\bigg|_{-\beta t}^t+\\
& +\frac{e^{-2i(\omega+\omega'-i\varepsilon)t}}{i(\omega+\omega'-i\varepsilon)}-\frac{1}{i(\omega+\omega'+i\varepsilon)}=\frac{e^{-i(\omega+\omega'-i\varepsilon)(1-\beta)t}}{i(\omega+\omega'-i\varepsilon)}+\frac{\alpha-1}{i(\omega+\omega'+i\varepsilon)}-\alpha\frac{e^{-i(\omega+\omega'+i\varepsilon)(1-\beta)t}}{i(\omega+\omega'+i\varepsilon)}.
\end{split} \end{equation}
Consequently, after simplifying the Hamiltonian, we have:
\begin{equation}
\begin{split}
& H= \theta(-t)H^{0}+\theta(t)\iint \frac{d\omega}{2\pi}\frac{d\omega'}{2\pi}\frac{\sqrt{\omega\omega'}}{2}\operatorname{v.p.}\frac{1}{\omega+\omega'}\frac{1}{i} \big[(1-\alpha) e^{-i(\omega+\omega')(1-\beta)t}+(\alpha-1)\big]a_\omega a_{\omega'}+\\
& +\theta(t)\iint \frac{d\omega}{2\pi}\frac{d\omega'}{2\pi}\frac{\sqrt{\omega\omega'}}{2}\big\{\omega\to-\omega,\ \omega'\to-\omega'\big\}a_\omega^{\dag} a_{\omega'}^{\dag}+\\
& +\theta(t)\iint \frac{d\omega}{2\pi}\frac{d\omega'}{2\pi}\frac{\sqrt{\omega\omega'}}{2} \operatorname{v.p.}\frac{1}{\omega-\omega'}\frac{1}{i}\big[(1-\alpha) e^{-i(\omega-\omega')(1-\beta)t}+(\alpha-1)\big]a_\omega a_{\omega'}^{\dag}+\\
& +\theta(t)\iint \frac{d\omega}{2\pi}\frac{d\omega'}{2\pi}\frac{\sqrt{\omega\omega'}}{2}\big\{\omega\to-\omega,\ \omega'\to-\omega'\big\}a_\omega^{\dag} a_{\omega'}+\theta(t) \frac{1}{2}\int\frac{d\omega}{2\pi}\omega\big[a_\omega^{\dag}a_\omega+a_\omega a_\omega^{\dag}\big].
\end{split}
\end{equation}
Performing the same steps as above, we also obtain the momentum operator:
\begin{equation}\begin{split}& P=\theta(-t)P^{0}+\theta(t)\iint \frac{d\omega}{2\pi}\frac{d\omega'}{2\pi}\frac{\sqrt{\omega\omega'}}{2}\operatorname{v.p.}\frac{1}{\omega+\omega'}\frac{1}{i} \big[e^{-i(\omega+\omega')(1-\beta)t}(1+\alpha)+(1-\alpha)\big]a_\omega a_{\omega'}+\\
& +\theta(t)\iint \frac{d\omega}{2\pi}\frac{d\omega'}{2\pi}\frac{\sqrt{\omega\omega'}}{2}\big\{\omega\to-\omega,\ \omega'\to-\omega'\big\}a_\omega^{\dag} a_{\omega'}^{\dag}+\\
&+\theta(t)\iint \frac{d\omega}{2\pi}\frac{d\omega'}{2\pi}\frac{\sqrt{\omega\omega'}}{2} \operatorname{v.p.}\frac{1}{\omega-\omega'}\frac{1}{i} \big[e^{-i(\omega-\omega')(1-\beta)t}(1+\alpha)+(1-\alpha)\big]a_\omega a_{\omega'}^{\dag}+\\
& +\theta(t)\iint \frac{d\omega}{2\pi}\frac{d\omega'}{2\pi}\frac{\sqrt{\omega\omega'}}{2}\big\{\omega\to-\omega,\ \omega'\to-\omega'\big\}a_\omega^{\dag} a_{\omega'}.\end{split}\end{equation}
One can see that the coefficients in the expression of the Hamiltonian and of the momentum operator through creation and annihilation operators depend on time. However, if we now calculate the linear combination $H-\beta P$, this dependence on time disappears. Namely,
\begin{equation} \label{eq:lincomb1}
\begin{split}
& H-\beta(t) P  =\theta(-t)H^0+\theta(t)\iint\frac{d\omega}{2\pi}\frac{d\omega'}{2\pi}\frac{\sqrt{\omega\omega'}}{2i}(\alpha-1)(1+\beta)\bigg[\operatorname{v.p.}\frac{1}{\omega+\omega'}\ a_\omega a_{\omega'}+ \\
& +\operatorname{v.p}\frac{1}{-\omega-\omega'}\ a_\omega^{\dag} a_{\omega'}^{\dag}+\operatorname{v.p}\frac{1}{\omega-\omega'}\ a_\omega a_{\omega'}^\dag+\operatorname{v.p}\frac{1}{-\omega+\omega'}\ a_\omega^{\dag} a_{\omega'}\bigg]+\theta(t) \frac{1}{2}\int\frac{d\omega}{2\pi}\omega\big[a_\omega^{\dag}a_\omega+a_\omega a_\omega^{\dag}\big],
\end{split}
\end{equation}
where we use that in the case under study $\beta = 0$ for $t<0$ and $\beta = const \neq 0$ for $t>0$.

Thus, after $t=0$, the linear combination $H - \beta(t)\, P$ can be diagonalized. Furthermore, we already know mode functions and creation-annihilation operators, diagonalizing it in the future, namely, those we obtained in the previous subsection.\\

\section{General expressions for loop corrections} \label{GeneralExpForLoops}
\quad\quad Let us now turn to the calculation of loop corrections due to the interaction $\lambda\phi^4$ with the use of Schwinger-Keldysh diagram technique (one can find its derivation in \cite{LL},\cite{Kamenev}). Two-loop corrections to all propagators $G^{\pm \pm}_{xy},\ G^{\pm \mp}_{xy}$ have the following form (up to a numerical factor) (for more details see, for example, \cite{AkhGodPop},\cite{LL}):

\begin{equation}\Delta G_{xy}=\lambda^2 \int dz\  dw\  G_{xz} \Sigma_{zw} G_{wy},\end{equation}
where

$$G=\begin{bmatrix}
     G^{--}      & G^{-+} \\
     G^{+-}     & G^{++}
\end{bmatrix}\
\operatorname{and}\
\Sigma=\begin{bmatrix}
    - G^{--3}      & \ \  G^{-+3} \\
    \ \  G^{+-3}     & -G^{++3}
\end{bmatrix}.
$$
In particular,
$$\Delta G^{--}_{xy}=\lambda^2\int d^2z\  d^2w \bigg[ -G^{--}_{xz}G^{--3}_{zw} G^{--}_{wy}+G^{--}_{xz}G^{-+3}_{zw}G^{+-}_{wy}+G^{-+}_{xz}G^{+-3}_{zw}G^{--}_{wy}-G^{-+}_{xz}G^{++3}_{zw}G^{+-}_{wy}\bigg],$$
$$\Delta G^{++}_{xy}=\lambda^2\int d^2z\  d^2w \bigg[-G^{+-}_{xz}G^{--3}_{zw} G^{-+}_{wy}+G^{+-}_{xz}G^{-+3}_{zw}G^{++}_{wy}+G^{++}_{xz}G^{+-3}_{zw}G^{-+}_{wy}-G^{++}_{xz}G^{++3}_{zw}G^{++}_{wy}\bigg],$$
and
\begin{equation}
\label{eq:keld1}
\Delta G^{K}_{xy}=\frac{1}{2}\big[\Delta G^{--}_{xy}+\Delta G^{++}_{xy}\big].\end{equation}
This correction to the Keldysh propagator is the sum of the following diagrams:

\begin{equation}
\label{eq:diag}
\Delta G^{K}_{xy}=\sum_{\sigma,\sigma_i=\pm1} \begin{gathered}
\begin{tikzpicture}

\begin{feynman}

 \vertex[label=${x, \sigma}$] (x);
    \vertex[label=${z, \sigma_1 }\ \ \ \ \ \ $] [right=2.5cm of x] (z);
    \vertex[label=$\ \ \ \ \ \ \ {w, \sigma_2}$] [right=5cm of z] (w) ;
    \vertex[label=${y, \sigma}$] [right=2.5cm of w] (y);

    \diagram* {

      (x) -- [plain] (z) -- [plain] (w) -- [plain] (y)
      (z) -- [plain, out=80, in=100] (w)
      (z) -- [plain, out=-80, in=-100] (w)

    };

   \end{feynman}
 \end{tikzpicture}
    \end{gathered}
    \end{equation}
Let us write out expressions for propagators $G^{\sigma_1 \sigma_2}_{xy}$ through the modes $h_k$:
\begin{equation}
\begin{split}
\label{eq:throughmodes}
& G^{--}_{xy}=\braket{T\phi(x)\phi(y)}=\theta(x^0-y^0)\int \frac{dk}{2\pi}\ h_k(x)\bar{h}_k(y)+\theta(y^0-x^0)\int \frac{dk}{2\pi}\ h_k(y)\bar{h}_k(x); \\
& G^{++}_{xy}=\braket{\tilde{T}\phi(x)\phi(y)}=\theta(x^0-y^0)\int \frac{dk}{2\pi}\ h_k(y)\bar{h}_k(x)+\theta(y^0-x^0)\int \frac{dk}{2\pi}\ h_k(x)\bar{h}_k(y); \\
& G^{+-}_{xy}=\braket{\phi(x)\phi(y)}=\int \frac{dk}{2\pi}\ h_k(x)\bar{h}_k(y);\ G^{-+}_{xy}=\braket{\phi(y)\phi(x)}=G^{+-}_{yx}.
\end{split}
\end{equation}
Note that due to $G^{--}_{x_1x_2}=\big(G^{++}_{x_1x_2}\big)^*$ and $G^{-+}_{x_1x_2}=\big(G^{+-}_{x_1x_2}\big)^*$, we have that $\Delta G^{--}_{xy}=\big(\Delta G^{++}_{xy}\big)^*$. Substituting expressions of propagators through the modes into eq. (\ref{eq:keld1}), we see that the correction to Keldysh propagator can be rewritten as follows:
\begin{equation}
\label{eq:occupanomal}
\Delta G^{K}_{xy}=\iint \frac{dk}{2\pi}\frac{dk'}{2\pi} \big[n_{kk'} \bar{h}_k(x) h_{k'}(y)+\kappa_{kk'} h_k(x)h_{k'}(y)+\operatorname{h.c.}\big],\end{equation}
where:
\begin{equation}\begin{split}n_{kk'}= & \lambda^2\int d^2z\  d^2w\  {h}_k(z) \bar{h}_{k'}(w) \bigg[\int \frac{dp}{2\pi} h_p(z) \bar{h}_p(w)\bigg]^3\big[-\theta(z^0-x^0)\theta(z^0-w^0)\theta(y^0-w^0)+\theta(y^0-w^0)+\\
& +\theta(x^0-z^0)-\theta(x^0-z^0)\theta(w^0-z^0)\theta(w^0-y^0)\big]+\\
& +\lambda^2\int d^2z\  d^2w\  {h}_k(z) \bar{h}_{k'}(w) \bigg[\int \frac{dp}{2\pi} h_p(w) \bar{h}_p(z)\bigg]^3\big[-\theta(z^0-x^0)\theta(w^0-z^0)\theta(y^0-w^0)-\\
& -\theta(x^0-z^0)\theta(z^0-w^0)\theta(w^0-y^0)\big].\end{split}\end{equation}
In the limit we are interested in (namely, $(x^0+y^0)/2\to \infty$, while $|x^0-y^0|=\operatorname{const}$), one may put $x^0\approx y^0\approx T\to \infty$ in the leading approximation. Substituting $x^0\approx y^0\approx T$ into the formula for $n_{kk'}$ and simplifying factors, containing theta-functions, we obtain the leading expression:

\begin{equation}
\label{eq:nkk}
n_{kk'} (T) \approx 2\lambda^2\int d^2z\  d^2w\ \theta(T-w^0)\theta(T-z^0)\  {h}_k(z) \bar{h}_{k'}(w) \bigg[\int \frac{dp}{2\pi} h_p(z) \bar{h}_p(w)\bigg]^3.\end{equation}
Similarly:
\begin{equation}\begin{split} & \kappa_{kk'}= \lambda^2\int d^2z\  d^2w\  \bar{h}_k(z) \bar{h}_{k'}(w) \bigg[\int \frac{dp}{2\pi} h_p(z) \bar{h}_p(w)\bigg]^3\big[-\theta(x^0-z^0)\theta(z^0-w^0)\theta(y^0-w^0)-\theta(z^0-w^0)+\\
& +\theta(z^0-x^0)-\theta(z^0-x^0)\theta(w^0-z^0)\theta(w^0-y^0) \big]+\\
& +\lambda^2\int d^2z\  d^2w\  \bar{h}_k(z) \bar{h}_{k'}(w) \bigg[\int \frac{dp}{2\pi} h_p(w) \bar{h}_p(z)\bigg]^3\big[-\theta(x^0-z^0)\theta(w^0-z^0)\theta(y^0-w^0)-\theta(w^0-z^0)+\\
& +\theta(w^0-y^0)-\theta(z^0-x^0)\theta(z^0-w^0)\theta(w^0-y^0) \big].\end{split}\end{equation}
After the substitution $x^0\approx y^0\approx T$, we obtain the leading expression in the limit $T\to +\infty$:
$$\kappa_{kk'}(T) \approx \lambda^2\int d^2z\  d^2w\  \bar{h}_k(z) \bar{h}_{k'}(w)\bigg\{-2\theta(T-z^0)\theta(z^0-w^0) \bigg[\int \frac{dp}{2\pi} h_p(z) \bar{h}_p(w)\bigg]^3-$$
$$-2\theta(T-w^0)\theta(w^0-z^0) \bigg[\int \frac{dp}{2\pi} h_p(w) \bar{h}_p(z)\bigg]^3\bigg\}.$$
Swapping the integration variables $w$ and $z$ in the second integral, we obtain the final expression:
\begin{equation}
\label{eq:kappakk}
\kappa_{kk'}(T) \approx -2\lambda^2\int d^2z\  d^2w\ \theta(T-z^0)\theta(z^0-w^0)\   \bigg[\bar{h}_k(z) \bar{h}_{k'}(w)+\bar{h}_k(w) \bar{h}_{k'}(z)\bigg] \bigg[\int \frac{dp}{2\pi} h_p(z) \bar{h}_p(w)\bigg]^3.\end{equation}
Below we calculate the explicit behaviour of $n_{kk'}(T)$ and $\kappa_{kk'}(T)$ as $T\to \infty$, via substitution into the above equations different concrete forms of modes $h_k$ for various types of mirror motions.

\section{Calculation of two-loop corrections to $n_{kk'}$} \label{twoloopnkk} \

In this section we calculate $n_{kk'}$ (\ref{eq:nkk}) for various types of mirror world-lines.

\subsection{The situation in the empty space--time (without mirror)} \label{nomirror}\

Let us first consider the simplest case, when there is no mirror and we deal with the standard scalar field theory in the entire $1+1$ Minkowski space-time. For this calculation we keep the mass $m$ arbitrary. Modes in this case are plane waves:
$$h_k(t,x)=\frac{1}{\sqrt{2\varepsilon_k}}e^{-i(\varepsilon_k t-kx)},\  \operatorname{where}\ \varepsilon_k=\sqrt{k^2+m^2}\ \operatorname{and}\ -\infty<k<\infty.$$
In such a case the integral over the coordinates $z=(z^0, z^1)$ and $w=(w^0, w^1)$ in eq. (\ref{eq:nkk}) can be rewritten as follows:
\begin{equation}\int_{t_0}^T dz^0\ e^{-i(\varepsilon_k+\sum_i \varepsilon_{p_i})z^0} \int_{t_0}^T dw^0\   e^{i(\varepsilon_{k'}+\sum_i \varepsilon_{p_i})w^0}\int_{-\infty}^{\infty} dz^1\ e^{i(k+\sum_i p_i)z^1} \int_{-\infty}^{\infty} dw^1\ e^{-i(k'+\sum_i p_i)w^1}.\end{equation}
Obviously, in the limit $T\to \infty$, $t_0 \to -\infty$ this expression is equal to
\begin{equation}\begin{split} & \delta\bigg(\varepsilon_k+\sum_i \varepsilon_{p_i}\bigg)\delta\bigg(\varepsilon_{k'}+\sum_i \varepsilon_{p_i}\bigg)\delta\bigg(k+\sum_i p_i\bigg)\delta\bigg(k'+\sum_i p_i\bigg)=\\
&= \delta(0)\delta\bigg(\varepsilon_k+\sum_i \varepsilon_{p_i}\bigg)\delta\bigg(k+\sum_i p_i\bigg)\delta\big(k-k'\big) \propto\\
& \propto(T-t_0)\delta(k-k')\delta\bigg(\varepsilon_k+\sum_i \varepsilon_{p_i}\bigg)\delta\bigg(k+\sum_i p_i\bigg),\end{split}\end{equation}
where we have used that $\delta(0)\propto T-t_0$.\\

The result is proportional to $T-t_0$, but $(\varepsilon_k+\sum_i \varepsilon_{p_i})$ is never equal to $0$, so the correction to $n_{kk'}$ does not grow with time. In fact, it is equal to $0$. Note, however, that for the case when $m=0$ we may encounter the standard infrared divergences in the integrals defining $n_{kk'}$, which are coming from the mode normalisation factors. The same sort of divergences we will encounter below. To deal with them we will always assume that we consider the massive scalar field with $m\to 0$.

\subsection{The mirror at rest} \

Now let us consider the case, when there is the mirror and it is at rest. Calculating $n_{kk'}$, let us first take the integral over the variable $z$ (we use the modes (\ref{eq:2})):
\begin{equation}\begin{split} & \int d^2 z\  \theta(T-z^0)\  h_k(z) h_{p_1} (z) h_{p_2} (z) h_{p_3} (z)=\int_{t_0}^{T} dz^0 \int_{0}^{\infty} dz^1\ h_k(z) h_{p_1} (z) h_{p_2} (z) h_{p_3} (z) =\\
& =\frac{1}{\sqrt{2^4\cdot kp_1p_2p_3}} \int_{t_0}^{T} dz^0 \int_{0}^{\infty} dz^1\ (e^{-ikv}-e^{-iku})(e^{-ip_1v}-e^{-ip_1u})(e^{-ip_2v}-e^{-ip_2u})(e^{-ip_3v}-e^{-ip_3u})=\\
& =\frac{1}{\sqrt{2^4\cdot kp_1p_2p_3}} \int_{t_0}^{T} dz^0\  e^{-i(k+p_1+p_2+p_3)z^0} \int_{0}^{\infty} dz^1 \sum_{\sigma,\sigma_i=\pm 1} \sigma\sigma_1\sigma_2\sigma_3\  e^{-i(\sigma k +\sigma_1 p_1+\sigma_2 p_2 +\sigma_3 p_3)z^1}.
\end{split}
\end{equation}
There are $16$ terms in the sum, which are different from each other by the signs in front of $k, p_1, p_2, p_3$ in the argument of exponential function. If the number of "$-$"-signs  in the argument of exponent is even (odd), then this term comes with $+$ ($-$) in front of it.

The above expression is equal to:
\begin{equation*}\frac{1}{\sqrt{2^4\cdot kp_1p_2p_3}}\bigg(\sum_{\sigma,\sigma_i=\pm 1} \sigma\sigma_1\sigma_2\sigma_3\ \frac{1}{i(\sigma k+\sum_i \sigma_i p_i-i\epsilon)}\bigg) \int_{t_0}^{T} dz^0\  e^{-i(k+p_1+p_2+p_3)z^0}.\end{equation*}
When $T \to \infty$ and $t_0 \to -\infty$, the last integral tends to the $\delta$-function:
\begin{equation*}\int_{t_0}^{T} dz^0\  e^{-i(k+p_1+p_2+p_3)z^0} \approx 2\pi \delta (k+p_1+p_2+p_3).\end{equation*}
As $k,p_1,p_2,p_3\geq0$, the equality $k+p_1+p_2+p_3=0$ can be fulfilled only when $k=p_1=p_2=p_3=0$.

Therefore, \begin{equation*}\begin{split} & \int d^2 z\  \theta(T-z^0)\  h_k(z) h_{p_1} (z) h_{p_2} (z) h_{p_3} (z)=\frac{1}{\sqrt{2^4\cdot kp_1p_2p_3}}\cdot 2\pi\delta(k+p_1+p_2+p_3)\times \\
& \times\bigg(\sum_{\sigma,\sigma_i=\pm 1} \sigma\sigma_1\sigma_2\sigma_3\ \frac{1}{i(\sigma k+\sum_i \sigma_i p_i-i\epsilon)}\bigg). \end{split}\end{equation*}
Hence,
\begin{equation}
 \label{eq:nkk1}
 \begin{split}
 n_{kk'} (T)= &  \lambda^2 \frac{1}{2^4\cdot\sqrt{kk'}}\int \bigg(\prod_{i=1}^3\frac{dp_i}{2\pi}\bigg) \frac{1}{p_1p_2p_3}2\pi\delta(k+p_1+p_2+p_3)\cdot 2\pi\delta(k'+p_1+p_2+p_3)\times \\
& \times \bigg(\sum_{\sigma,\sigma_i=\pm 1} \sigma\sigma_1\sigma_2\sigma_3\ \frac{1}{i(\sigma k+\sum_i \sigma_i p_i-i\epsilon)}\bigg)\cdot \bigg(\sum_{\sigma,\sigma_i=\pm 1} \sigma\sigma_1\sigma_2\sigma_3\ \frac{-1}{i(\sigma k'+\sum_i \sigma_i p_i+i\epsilon)}\bigg).
  \end{split}
 \end{equation}
Note that due to the presence of the $\delta$-functions the expressions in the round brackets are equal to $0$, and $n_{kk'}=0$.\\

Above we again encounter the standard complication due to the infrared divergence of the massless case. It comes from the mode normalization factors at $p_{1,2,3} = k = k' = 0$. To understand what is going on here we consider
the massive case. In this case we should have the following integrand (compare with (\ref{eq:nkk1})):
\begin{equation}
\begin{split}
& \delta\bigg(\varepsilon_k+\sum_i \varepsilon_{p_i}\bigg)\delta\bigg(\varepsilon_{k'}+\sum_i \varepsilon_{p_i}\bigg)\cdot \bigg(\sum_{\sigma,\sigma_i=\pm 1} \sigma\sigma_1\sigma_2\sigma_3\ \frac{1}{i(\sigma k+\sum_i \sigma_i p_i-i\epsilon)}\bigg)\times \\
& \times \bigg(\sum_{\sigma,\sigma_i=\pm 1} \sigma\sigma_1\sigma_2\sigma_3\ \frac{-1}{i(\sigma k'+\sum_i \sigma_i p_i+i\epsilon)}\bigg).
\end{split}
\end{equation}
As in the section \ref{nomirror}, the argument of the $\delta$-function $\varepsilon_k+\sum_i \varepsilon_{p_i}$ is never equal to $0$, so the correction, in fact, is $0$. And we do not have here any infrared divergence.

\subsection{The broken world-line} \label{nkkbrokentraj} \

In this case we use the modes from eq. (\ref{eq:harmthet}). Then the integral over $z$ can be divided as follows:
$$\int d^2 z\  \theta(T-z^0)\  h_k(z) h_{p_1} (z) h_{p_2} (z) h_{p_3} (z)=\int_{t_0}^{0} dz^0 \int_{0}^{\infty} dz^1\ h_k^0(z) h_{p_1}^0 (z) h_{p_2}^0 (z) h_{p_3}^0 (z)+$$
$$+\int_{0}^{T} dz^0 \int_{-\beta z^0}^{z^0} dz^1\ h_k^\beta(z) h_{p_1}^\beta (z) h_{p_2}^\beta (z) h_{p_3}^\beta (z)+\int_{0}^{T} dz^0 \int_{z^0}^{\infty} dz^1\ h_k^0(z) h_{p_1}^0 (z) h_{p_2}^0 (z) h_{p_3}^0 (z).$$
Adding the term $$\int_0^T dz^0 \int_{0}^{z^0} dz^1\ h_k^0(z) h_{p_1}^0 (z) h_{p_2}^0 (z) h_{p_3}^0 (z) $$
to the r.h.s. of the equality above and subtracting it, we obtain the contribution to $n_{kk'}$ from the previous subsection: $$\int_{t_0}^{T} dz^0 \int_{0}^{\infty} dz^1\ h_k^0(z) h_{p_1}^0 (z) h_{p_2}^0 (z) h_{p_3}^0 (z),$$
which tends to $0$ when $T\to \infty,\ t_0 \to -\infty$. Consequently, in the limit $T\to \infty,\ t_0 \to -\infty$, the integral over $z$ can be simplified to:
$$\int d^2 z\  \theta(T-z^0)\  h_k(z) h_{p_1} (z) h_{p_2} (z) h_{p_3} (z)=\int_{0}^{T} dz^0 \int_{-\beta z^0}^{z^0} dz^1\ h_k^\beta(z) h_{p_1}^\beta (z) h_{p_2}^\beta (z) h_{p_3}^\beta (z)-$$
$$-\int_0^T dz^0 \int_{0}^{z^0} dz^1\ h_k^0(z) h_{p_1}^0 (z) h_{p_2}^0 (z) h_{p_3}^0 (z)=$$
\begin{equation}
\label{eq:simple}=\int_{0}^{T(1+\beta)} du \int_{\alpha u}^{2T-u} dv\ h_k^\beta h_{p_1}^\beta h_{p_2}^\beta h_{p_3}^\beta -\int_0^T du \int_{u}^{2T-u} dv\ h_k^0 h_{p_1}^0 h_{p_2}^0 h_{p_3}^0. \end{equation}
Let us calculate the first integral, neglecting exponentially small terms. We also omit the factor $1/\sqrt{2^4 kp_1p_2p_3}$, which is not important for the calculation of the integrals over $u$ and $v$ (we will put this factor back in the final result):
\begin{equation}\begin{split} & \int_{0}^{T(1+\beta)} du \int_{\alpha u}^{2T-u} dv\ h_k^\beta h_{p_1}^\beta h_{p_2}^\beta h_{p_3}^\beta= \int_{0}^{T(1+\beta)} du\  e^{-i(k+p_1+p_2+p_3-i\epsilon)\alpha u} \bigg[\frac{1}{i(k+p_1+p_2+p_3-i\epsilon)}+\\
& +2T-u(1+\alpha)-\frac{1}{i(k+p_1+p_2-i\epsilon)}-\frac{1}{i(p_3-i\epsilon)}+...\bigg].\end{split}\end{equation}
There are $15$ terms of the following form in the sum:
$$(-1)^{\sigma+\sigma_1+\sigma_2+\sigma_3}\  \frac{1}{i(\sigma k+ \sigma_1 p_1+ \sigma_2 p_2+ \sigma_3 p_3-i\epsilon)};\ \sigma, \sigma_1,\sigma_2, \sigma_3 = 0,1$$
(except for the case when $\sigma= \sigma_1=\sigma_3=\sigma_3 = 0$) and the term $2T-u(1+\alpha)$, which comes from the integration of the term $\displaystyle e^{-i\big(k+\sum_i p_i\big)}$ over $v$. Performing the integration over $u$, we obtain:
$$  \frac{1}{i\alpha(k+p_1+p_2+p_3-i\epsilon)} \bigg[\frac{1}{i(k+p_1+p_2+p_3-i\epsilon)}+2T+...\bigg]+\frac{1+\alpha}{\alpha^2} \frac{1}{(k+p_1+p_2+p_3-i\epsilon)^2}.$$
The last term comes from the integration of $u(1+\alpha)e^{-i(k+p_1+p_2+p_3-i\epsilon)\alpha u}$. Similarly, the second integral in eq. (\ref{eq:simple}) is equal to
$$  \frac{1}{i(k+p_1+p_2+p_3-i\epsilon)} \bigg[\frac{1}{i(k+p_1+p_2+p_3-i\epsilon)}+2T+...\bigg]+\frac{2}{(k+p_1+p_2+p_3-i\epsilon)^2}$$
(the sum in square brackets is the same as above; note that the second integral in eq. (\ref{eq:simple}) is equal to the first one, calculated at $\beta=0$).

\begin{figure}[h]
\centering
\includegraphics[scale=0.26]{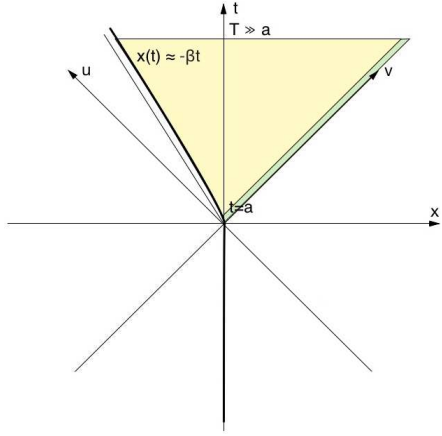}
\caption{The world-line (\ref{eq:asymptbeta}). In the yellow region the modes for the case of this world-line are approximately equal to those for the case of the broken world-line. In the green region the are different from each other.}
\label{fig:asymptbeta}
\end{figure}

Therefore, combining all these expressions together, we obtain that
\begin{equation}\begin{split} & \int d^2 z\  \theta(T-z^0)\  h_k(z) h_{p_1} (z) h_{p_2} (z) h_{p_3} (z)= \\
& = \frac{1-\alpha}{\alpha}\frac{1}{i(k+p_1+p_2+p_3-i\epsilon)} \bigg[\frac{1}{i(k+p_1+p_2+p_3-i\epsilon)}+2T-\frac{1}{i(k+p_1+p_2-i\epsilon)}-\frac{1}{i(p_3-i\epsilon)}+...\bigg]+\\
& +\bigg[\frac{1+\alpha}{\alpha^2}-2\bigg]\frac{1}{(k+p_1+p_2+p_3-i\epsilon)^2}.\end{split}\end{equation}
Now we are going to leave only the leading terms, which are growing with $T\to +\infty$. Such terms also appear from products of $\delta$-functions, such as $\delta(k+p_1+p_2+p_3)\delta(\sigma k+ \sigma_1 p_1+ \sigma_2 p_2+ \sigma_3 p_3)$   (where $\sigma, \sigma_1,\sigma_2, \sigma_3 = 0,1$), as they are equal to $\delta(k+p_1+p_2+p_3)\delta(0)=  \delta(k+p_1+p_2+p_3)\cdot (2T/\pi)$. Thus,\
\begin{equation}
\label{eq:halfnkkbroken}
\begin{split}
& \int d^2 z\  \theta(T-z^0)\  h_k(z) h_{p_1} (z) h_{p_2} (z) h_{p_3} (z)= \\
& =-\frac{2T(1-\alpha)}{\alpha}\bigg[ \operatorname{v.p.} \frac{i}{k+p_1+p_2+p_3}+\frac{1+2\alpha}{\alpha}\pi\delta(k+p_1+p_2+p_3)\bigg]+O(1).\end{split}
\end{equation}
Now we are ready to write out the final result for $n_{kk'}$ in the leading order in powers of $T$:
\begin{equation}
\label{eq:nkkbroken}
\begin{split}
 n_{kk'}\approx &\  2\lambda^2 (2T)^2\  \frac{(1-\alpha)^2}{\alpha^2} \int \prod_{j=1}^3 \frac{dp_j}{2\pi} \frac{1}{p_1p_2p_3} \frac{1}{2^4 \sqrt{kk'}}\bigg[\operatorname{v.p.} \frac{i}{k+p_1+p_2+p_3}+\frac{1+2\alpha}{\alpha}\pi\delta(k+p_1+p_2+p_3)\bigg]\times\\
& \times \bigg[-\operatorname{v.p.} \frac{i}{k'+p_1+p_2+p_3}+\frac{1+2\alpha}{\alpha}\pi\delta(k'+p_1+p_2+p_3)\bigg].
\end{split}\end{equation}
Similarly to the case of standing mirror the terms, containing $\delta$-functions give zero contribution, so the result can be simplified to:
\begin{equation}
\label{eq:nkkbrokensimple}
 n_{kk'}\approx \  2\lambda^2 (2T)^2\  \frac{(1-\alpha)^2}{\alpha^2} \int \prod_{j=1}^3 \frac{dp_j}{2\pi} \frac{1}{p_1p_2p_3} \frac{1}{2^4 \sqrt{kk'}}\bigg[\operatorname{v.p.} \frac{1}{k+p_1+p_2+p_3}\bigg]\cdot\bigg[ \operatorname{v.p.} \frac{1}{k'+p_1+p_2+p_3}\bigg].
\end{equation}
It is probably worth stressing at this point that in the limit $T\to \infty$ the loop corrections for the case of the broken world-line well approximate the corrections for the case of the world--line, which describes the eternal approach of the mirror to the with the constant speed $\beta<1$, as $t\to \infty$:
 \begin{equation}\label{eq:asymptbeta}x(t)=\begin{cases}
\ \ 0 & t<0\\

  -\beta t+a(1-e^{-\beta t/a})& t\ge0,
\end{cases}\end{equation}
In this case one should use the modes $e^{-ikv}-e^{-ikp(u)}$ instead of (\ref{eq:harmthet}), but they coincide with each other in the region $u<0$ and are almost equal in the region $u \gg a$. One should also integrate over $z^1$ from $x(z^0)$ rather than from $-\beta z^0$, but $x(t) \approx -\beta t$ when $t\gg u$ (see figure \ref{fig:asymptbeta}). So, in the limit $T\to \infty$ the loop corrections in the cases of the broken world-line and of the world-line (\ref{eq:asymptbeta}) are approximately equal.

\subsection{The mirror, approaching the speed of light} \label{nkkexp}

Let us calculate $n_{kk'}(T)$ for the mirror world-line (\ref{eq:exptraj}). In this the modes can be represented as:
\begin{equation}
\label{eq:harmexp}
h_k(u,v)=\theta(-u) h_k^0 (u,v)+\theta(u) h_k^\beta (u,v),\end{equation}
where $$h_k^0 (u,v)=\frac{i}{\sqrt{2k}}\big[e^{-ikv}-e^{-iku}\big],\quad {\rm and} \quad h_k^\beta (u,v)=\frac{i}{\sqrt{2k}}\big[e^{-ikv}-e^{-ik(2t_u-u)}\big].$$
Here $t_u$ is defined by the equation:
$$t_u-x(t_u)=u.$$
Therefore, for this particular world-line
$$2t_u-u=a(1-e^{-t_u/a}).$$
Note that any world-line $x(t)$, expressed in coordinates $(u,v)$, can be rewritten as follows:
\begin{equation}
\label{eq:vutraj}
v(u)=2t_u-u=t_u+x(t_u).\end{equation}
Applying the same arguments as in the previous subsection, in the limit $T\to \infty$, $t_0 \to -\infty$ we obtain the following expression for the integral over $z$ (which is a part of $n_{kk'}$ (\ref{eq:nkk})):
\begin{equation}\begin{split}
\label{eq:simpleexp} & \int d^2 z\  \theta(T-z^0)\  h_k(z) h_{p_1} (z) h_{p_2} (z) h_{p_3} (z)=\int_{0}^{T} dz^0 \int_{x(z^0)}^{z^0} dz^1\ h_k^\beta(z) h_{p_1}^\beta (z) h_{p_2}^\beta (z) h_{p_3}^\beta (z)-\\
& -\int_0^T dz^0 \int_{0}^{z^0} dz^1\ h_k^0(z) h_{p_1}^0 (z) h_{p_2}^0 (z) h_{p_3}^0 (z)=\\
& =\int_{0}^{T-x(T)} du \int_{2t_u-u}^{2T-u} dv\ h_k^\beta h_{p_1}^\beta h_{p_2}^\beta h_{p_3}^\beta -\int_0^T du \int_{u}^{2T-u} dv\ h_k^0 h_{p_1}^0 h_{p_2}^0 h_{p_3}^0. \end{split}\end{equation}
The second integral was calculated in the previous section. Let us compute the first one:
\begin{equation}\begin{split} & \int_{0}^{T-x(T)} du \int_{2t_u-u}^{2T-u} dv\ h_k^\beta h_{p_1}^\beta h_{p_2}^\beta h_{p_3}^\beta= \int_{0}^{T-x(T)} du\  e^{-i(k+p_1+p_2+p_3-i\epsilon)(2t_u-u)} \bigg[\frac{1}{i(k+p_1+p_2+p_3-i\epsilon)}+\\
& +2T-u-(2t_u-u)-\frac{1}{i(k+p_1+p_2-i\epsilon)}-\frac{1}{i(p_3-i\epsilon)}+...\bigg].\end{split}\end{equation}
In this sum there are $15$ terms of the following type:
$$(-1)^{\sigma+\sigma_1+\sigma_2+\sigma_3}\  \frac{1}{i(\sigma k+ \sigma_1 p_1+ \sigma_2 p_2+ \sigma_3 p_3-i\epsilon)};\ \sigma, \sigma_1,\sigma_2, \sigma_3 = 0,1$$
(except for the case when $\sigma= \sigma_1=\sigma_3=\sigma_3 = 0$) and the term $2T-u-(2t_u-u)=2(T-t_u)$. It is more convenient to change the variable $u$ to $t_u$ in the last integral, using the fact that $$dt_u\bigg[1-\frac{dx(t_u)}{dt_u}\bigg]=du.$$
Then the previous integral transforms into
\begin{equation}
\begin{split}
&\int_{0}^{T} dt_u\bigg[1-\frac{dx(t_u)}{dt_u}\bigg]\  e^{-i(k+p_1+p_2+p_3-i\epsilon)(t_u+x(t_u))} \bigg[\frac{1}{i(k+p_1+p_2+p_3-i\epsilon)}+\\
&+2T-2t_u-\frac{1}{i(k+p_1+p_2-i\epsilon)}-\frac{1}{i(p_3-i\epsilon)}+...\bigg].
\end{split}
\end{equation}
Note that this formula works for an arbitrary mirror world-line $x(t)$. Substituting the world-line under consideration into it, we obtain:
$$ \int_{0}^{T} dt_u\bigg[2-e^{-t_u/a}\bigg]\  e^{-i(k+p_1+p_2+p_3-i\epsilon)a(1-e^{-t_u/a})} \bigg[\frac{1}{i(k+p_1+p_2+p_3-i\epsilon)}+$$
$$+2T-2t_u-\frac{1}{i(k+p_1+p_2-i\epsilon)}-\frac{1}{i(p_3-i\epsilon)}+...\bigg].$$
It is probably worth stressing here that if the world-line had the form $x(t)=-\beta t+a(1-e^{-\beta t/a})$ with $\beta < 1$ at $t>0$, then as a result of the integration we would obtain $\delta$-functions due to proportionality of the argument of the exponent to $t_u$. Indeed, in such a case we would have had $t_u+x(t_u)=t_u(1-\beta)+a(1-e^{-\beta t/a})$. When $\beta = 1$, however, the linear term in $t_u$ vanishes and the situation is different.\\

To continue we write out all the integrals from the previous formula:
\begin{equation*}
\begin{split}
& \int_0^T dt_u\ e^{-t_u/a}\ e^{-i(k-i\epsilon)a e^{-t_u/a}}=\frac{e^{i(k-i\epsilon)a}-e^{i(k-i\epsilon)ae^{-T/a}}}{i(k-i\epsilon)}\to \frac{e^{i(k-i\epsilon)a}-1}{i(k-i\epsilon)},\ T\to \infty\\
& \int_0^T dt_u\ e^{-i(k-i\epsilon)a e^{-t_u/a}}=T+\sum_{n=1}^{\infty} \frac{a^{n+1}i^n(k-i\epsilon)^n}{n\cdot n!},\ T\to \infty\\
& \int_0^T dt_u\ t_u e^{-t_u/a}\cdot e^{-i(k-i\epsilon)a e^{-t_u/a}}=\sum_{n=1}^{\infty} \frac{(-1)^n(ia)^{n+1}(k-i\epsilon)^{n-1}}{n\cdot n!},\ T\to \infty\\
& \int_0^T dt_u\ t_u e^{-i(k-i\epsilon)a e^{-t_u/a}}=\frac{T^2}{2}+ ia\sum_{n=1}^{\infty}\int_{0}^{k-i\epsilon} dp\  \frac{(-1)^n (ia)^{n+1}p^{n-1}}{n\cdot n!},\ T\to \infty
\end{split}
\end{equation*}
In the formulae above we neglected exponentially small terms ($e^{-T/a}\to 0$). We would like to keep track only of the leading terms as $T\to \infty$. What is important is that some of these integrals give a power of $T$ and that all powers of $(k-i\epsilon)$ are non-negative. Thus, because of the behaviour of the world-line $z(t)$ at $t\to \infty$, integration over $u$ does not yield $\delta$-functions , but it does yield powers of $T$. Now we are ready to write the complete expression for the integral over $z$:
$$\int d^2 z\  \theta(T-z^0)\  h_k(z) h_{p_1} (z) h_{p_2} (z) h_{p_3} (z)= \bigg[\frac{1}{i(k+p_1+p_2+p_3-i\epsilon)}+$$
$$+2T-\frac{1}{i(k+p_1+p_2-i\epsilon)}-\frac{1}{i(p_3-i\epsilon)}+...\bigg]\bigg[e^{-i(k+p_1+p_2+p_3-i\epsilon)a}\bigg(2T+\sum_{n=1}^{\infty}\operatorname{...}\bigg)-\frac{1}{i(k+p_1+p_2+p_3-i\epsilon)}\bigg]-$$
$$-2e^{-i(k+p_1+p_2+p_3-i\epsilon)a}\bigg[-\frac{T^2}{2}+\sum_{n=1}^{\infty}\operatorname{...}\bigg]-\frac{2}{(k+p_1+p_2+p_3-i\epsilon)^2},$$
where dots are standing for the subleading terms. We see that in this case, the leading term is of the order of $T^2$:

\begin{equation}
\label{eq:halfexp}
\int d^2 z\  \theta(T-z^0)\  h_k(z) h_{p_1} (z) h_{p_2} (z) h_{p_3} (z)=5T^2 e^{-i(k+p_1+p_2+p_3-i\epsilon)a}+O(T).\end{equation}
Finally, the expression for $n_{kk'}$ to the leading order in powers of $T$ (up to a numerical coefficient) is as follows:
\begin{equation}
\label{eq:nkkexp}
n_{kk'}\propto \lambda^2 T^4\  \frac{e^{-i(k-k'-i\epsilon)a}}{ \sqrt{kk'}}\cdot  \int \prod_{j=1}^3 \frac{dp_j}{2\pi} \frac{1}{p_1p_2p_3}.
\end{equation}
\begin{figure}[h]
\centering
\includegraphics[scale=0.26]{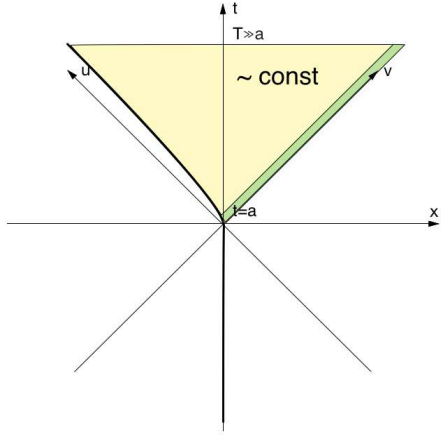}
\caption{The region of integration in the integral (\ref{eq:eqeq}). In the yellow region the reflected wave is almost constant. In the green region it behaves like a plane wave.}
\label{fig:6}
\end{figure}

This result can be easily understood. To calculate $n_{kk'}$, one has to take the following integral over $z$:
\begin{equation}
\label{eq:eqeq}\int_{0}^{T} dz^0 \int_{x(z^0)}^{z^0} dz\ h_k^\beta(z) h_{p_1}^\beta (z) h_{p_2}^\beta (z) h_{p_3}^\beta (z),\end{equation}
where $h_p^\beta(z)\propto e^{-ipv}-e^{-ipa(1-e^{-t_u/a})}$. The reflected wave behaves like a constant in the region $u\gg a$, and like a plane wave in the region $u\ll a$:
$$ e^{-ipa(1-e^{-t_u/a})} \propto \begin{cases}
e^{-ipa}\ \ \ u\gg a\\
e^{-ipt_u}\ \   u\ll a.
\end{cases}
$$
As the result, from the integration over $z^0, z$ in the limit when $T\gg a$, one obtains something approximately proportional to the volume of the region of integration $\propto T^2$ due to the behaviour of reflected waves (see figure \ref{fig:6}). Thus, $n_{kk'}$ is proportional to $T^2 \cdot T^2=T^4$.\\

However, the integral over $p_i$ in eq. (\ref{eq:nkkexp}) contains ultraviolet divergence\footnote{They are present on top of the infrared divergences, which are typical in such integrals for the massless field and can be regularized by introducing a small mass, as we have explained above.}. To regulate these divergences one has to consider a non-perfect mirror of some sort. In fact, the divergences come from the fact that the mode functions in this case behave as constants when $u \gg a$. In the case of non--perfect mirror the mode functions will depend on space and time coordinates in the limit in question. This dependence should regulate the divergence under consideration. Just to give an idea how it should work, consider modes, which have the following from

$$
h_k(u,v) \propto \alpha(k) \, e^{- i \, k \, v} - \beta(k) \, e^{- i \, k \, p(u)}.
$$
In the presence of a non--ideal mirror modes should have some similar form. In fact, imagine that $\alpha(k) \sim \beta(k) \sim 1$ for $|k| < \Lambda$ and $\beta(k) = 0$ for $|k| > \Lambda$, where $\Lambda$ is some energy scale. The behavior of the modes under consideration just means that there is a mirror which reflects all the modes with $|k| < \Lambda$ and is transparent for those with $|k| > \Lambda$. If one recalculates the loop corrections with the use of such modes, then instead of (\ref{eq:nkkexp}) he will obtain:

\begin{equation}
\label{eq:modified}
n_{kk'}\propto \lambda^2 T^4\, \beta(k) \bar{\beta}(k')  \frac{e^{-i(k-k'-i\epsilon)a}}{ \sqrt{kk'}}\cdot  \int \prod_{j=1}^3 \frac{dp_j \, \left|\beta(p_j)\right|^2}{2\pi p_j},
\end{equation}
which obviously does not have any UV divergence. But, in any case, a more rigorous calculation will be done elsewhere.
\section{Calculation of two-loop corrections to $\kappa_{kk'}$} \label{twoloopkappa} \
In this section we calculate $\kappa_{kk'}$ (\ref{eq:kappakk}) for the same mirror world-lines as have been considered above.

\subsection{The mirror at rest}\

Now we calculate the following quantity, which is the part of $\kappa_{kk'}$:
\begin{equation}
\label{eq:partkappa}
\int \bigg(\prod_{i=1}^3\frac{dp_i}{2\pi}\bigg) \int_{t_0}^T dz^0 \int_{0}^{\infty} dz^1\  \bar h_k(z^0,z) \prod_{i=1}^3 h_{p_i}(z^0,z) \int_{t_0}^{z^0} dw^0 \int_{0}^\infty dw^1\  \bar h_{k'}(w^0,w)\prod_{i=1}^3 \bar h_{p_i} (w^0,w).\end{equation}
Integrating over $z$ and $w$, we obtain:
\begin{equation}\begin{split}& -\frac{\pi^2}{\sqrt{kk'}}\int \bigg(\prod_{i=1}^3\frac{dp_i}{2\pi}\bigg)\  \frac{1}{2^4p_1p_2p_3} \int_{t_0}^T dz^0\  e^{-i(-k+p_1+p_2+p_3)z^0} \int_{t_0}^{z^0} dw^0\  e^{i(k'+p_1+p_2+p_3)w^0}\times\\
& \times \bigg[\sum_{\sigma,\sigma_i=\pm1} \sigma\sigma_1\sigma_2\sigma_3\  \delta (-\sigma k+\sigma_1 p_1+\sigma_2 p_2+\sigma_3 p_3)\bigg]\cdot \bigg[\sum_{\sigma,\sigma_i=\pm1} \sigma\sigma_1\sigma_2\sigma_3\  \delta (\sigma k'+\sigma_1 p_1+\sigma_2 p_2+\sigma_3 p_3)\bigg].\end{split}\end{equation}
Taking the limit $T\to \infty, t_0 \to -\infty$, we integrate over $z^0$ and $w^0$:
\begin{equation}\begin{split} & \int_{-\infty}^{\infty} dz^0\  e^{-i(-k+p_1+p_2+p_3)z^0} \int_{-\infty}^{z^0} dw^0\  e^{i(k'+p_1+p_2+p_3)w^0}=\bigg|w^0+z^0=x_{+};\ w^0-z^0=x_{-}\bigg|=\\
&=\int_{-\infty}^0 dx_{-}\ e^{i(p_1+p_2+p_3+k'/2-k/2)x_{-}} \int_{-\infty}^{\infty} dx_{+}\ e^{i(k+k')x_{+}/2}=4\pi\delta(k+k')\cdot\frac{1}{i(p_1+p_2+p_3+(k'-k)/2-i\epsilon)}.\end{split}\end{equation}
This function differs from $0$ only when $k=k'=0$, which means that the correction to $\kappa_{kk'}$ can be rewritten as follows:
\begin{equation}
\label{eq:kapparest}
\begin{split}
\kappa_{kk'}=&-\frac{\pi^2}{\sqrt{kk'}}\int \bigg(\prod_{i=1}^3\frac{dp_i}{2\pi}\bigg)\  \frac{1}{2^4p_1p_2p_3} \cdot\frac{1}{i(p_1+p_2+p_3+(k'-k)/2-i\epsilon)}\cdot4\pi\delta(k+k')\times\\
&\times \bigg[\sum_{\sigma,\sigma_i=\pm1} \sigma\sigma_1\sigma_2\sigma_3\  \delta (\sigma_1 p_1+\sigma_2 p_2+\sigma_3 p_3)\bigg]^2.\end{split}\end{equation}\
Obviously, the sum in square brackets is equal to $0$. As is expected, in this case $\kappa_{kk'}=0$.

\subsection{The broken world-line} \

The integral over $(z^0,z^1)$ and $(w^0,w^1)$ in (\ref{eq:partkappa}) can be represented as follows:
$$\bigg[\int_{0}^{T-x(T)} du_z \int_{2t_{u_z}-u_z}^{2T-u_z} dv_z\ \bar h_{k}^\beta(z)\prod_{i=1}^3 h_{p_i}^\beta (z)+\int_{-\infty}^{0} du_z \int_{u_z}^{2T-u_z} dv_z\ \bar h_{k}^0(z)\prod_{i=1}^3 h_{p_i}^0 (z)\bigg]\times$$
\begin{equation}
\label{eq:generalf}
\times\bigg[\int_{0}^{z^0-x(z^0)} du_w \int_{2t_{u_w}-u_w}^{2z^0-u_w} dv_w\ \bar h_{k'}^\beta(w)\prod_{i=1}^3 \bar h_{p_i}^\beta (w)+\int_{-\infty}^{0} du_w \int_{u_w}^{2z^0-u_w} dv_w\ \bar h_{k'}^0(w)\prod_{i=1}^3 \bar h_{p_i}^0 (w)\bigg],\end{equation}
where $\displaystyle z^0=\frac{u_z+v_z}{2}$. This equation ia applicable to any mirror world-line $x(t)$.  In the case of the broken world-line the integrals over $u_w$ and $v_w$ in eq. (\ref{eq:generalf}) turn into

$$\int_{0}^{z^0(1+\beta)} du_w \int_{\alpha u_w}^{2z^0-u_w} dv_w\ \bar h_{k'}^\beta(w)\prod_{i=1}^3 \bar h_{p_i}^\beta (w)=\int_{0}^{z^0(1+\beta)} du_w\ \bigg \{e^{i(k'+p_1+p_2+p_3)\alpha u_w}(2z^0-u_w-\alpha u_w)+$$
$$+ \frac{e^{i(k'+p_1+p_2+p_3)(2z^0-u_w)}}{i(k'+p_1+p_2+p_3)}-\frac{e^{ip_3(2z^0-u_w)}e^{i(k'+p_1+p_2)\alpha u_w}}{ip_3}+\frac{e^{i(k'+p_3)(2z^0-u_w)}e^{i(p_1+p_2)\alpha u_w}}{i(k'+p_3)}+...-$$
\begin{equation}
\label{eq:zhibroken1}
-e^{i(k'+p_1+p_2+p_3)\alpha u_w}\bigg[\frac{1}{i(k'+p_1+p_2+p_3)}-\frac{1}{ip_3}+\frac{1}{i(k'+p_3)}+...\bigg]\bigg\}.\end{equation}
The ellipsis denote the remaining contributions, which appear after the integration over $v_w$ of the following expressions (with $\sigma,\sigma_i=0,1$):

$$e^{i(\sigma k'+\sigma_1p_1+\sigma_2p_2+\sigma_3p_3)v_w}e^{i[ (1-\sigma) k'+(1-\sigma_1)p_1+(1-\sigma_2)p_2+(1-\sigma_3)p_3]\alpha u_w}.$$
An integral of an analytic function over a finite region is again an analytic function. So, we do not need any regularization in the expression above. It can be directly checked that when $k',p_i \to 0$, all singularities disappear. The first term in (\ref{eq:zhibroken1}) will give the leading contribution to $\kappa_{kk'}$ as $T\to\infty$ (the rest will give something of a smaller order in powers of $T$):
$$\int_{0}^{z^0(1+\beta)} du_w\ e^{i(k'+p_1+p_2+p_3)\alpha u_w}(2z^0-u_w-\alpha u_w)=\frac{-2z^0}{i(k'+\sum_i p_i)\alpha}-(1+\alpha)\frac{e^{i(k'+\sum_i p_i)z^0(1-\beta)}}{(k'+\sum_i p_i)^2\alpha^2}.$$
Let us now calculate the integral over $u_w$ and $v_w$ in the expression (\ref{eq:generalf}):
$$\int_{-\infty}^{0} du_w \int_{u_w}^{2z^0-u_w} dv_w\ \bar h_{k'}^0(w)\prod_{i=1}^3 \bar h_{p_i}^0 (w)=\int_{-\infty}^{0} du_w\ \bigg \{e^{i(k'+p_1+p_2+p_3) u_w}(2z^0-2u_w)+$$
$$+ \frac{e^{i(k'+p_1+p_2+p_3)(2z^0-u_w)}}{i(k'+p_1+p_2+p_3)}-\frac{e^{ip_3(2z^0-u_w)}e^{i(k'+p_1+p_2) u_w}}{ip_3}+\frac{e^{i(k'+p_3)(2z^0-u_w)}e^{i(p_1+p_2) u_w}}{i(k'+p_3)}+...-$$
\begin{equation}
\label{eq:zhibroken2}
-e^{i(k'+p_1+p_2+p_3) u_w}\bigg[\frac{1}{i(k'+p_1+p_2+p_3)}-\frac{1}{ip_3}+\frac{1}{i(k'+p_3)}+...\bigg]\bigg\}.\end{equation}
The ellipsis denote the remaining terms, appearing after the integration over $v_w$ of the following expressions (with $\sigma,\sigma_i=0,1$):

$$e^{i(\sigma k'+\sigma_1p_1+\sigma_2p_2+\sigma_3p_3)v_w}e^{i[ (1-\sigma) k'+(1-\sigma_1)p_1+(1-\sigma_2)p_2+(1-\sigma_3)p_3] u_w}.$$
The first term in (\ref{eq:zhibroken2}) will give the leading contribution to $\kappa_{kk'}$:

\begin{equation}\begin{split} & \int_{-\infty}^{0} du_w\ e^{i(k'+p_1+p_2+p_3) u_w}(2z^0-2u_w)=\frac{2z^0}{i(k'+p_1+p_2+p_3-i\epsilon)}-\frac{2}{(k'+p_1+p_2+p_3-i\epsilon)^2}\sim\\
& \sim\frac{2z^0}{i(k'+p_1+p_2+p_3-i\epsilon)}+2\pi^2\big[\delta(k'+p_1+p_2+p_3)\big]^2,\end{split}\end{equation}
with the last term giving $\delta(0)\delta(k'+p_1+p_2+p_3)\propto T\cdot\delta(k'+p_1+p_2+p_3).$ "$\sim$"-sign means that we discard the terms, which are negligibly small compared to what we keep here. So, the correction to $\kappa_{kk'}$ in the leading approximation looks as follows:
\begin{equation}\begin{split}\label{eq:leading1} & \bigg[\int_{0}^{T(1+\beta)} du_z \int_{\alpha u_z}^{2T-u_z} dv_z\ \bar h_{k}^\beta(z)\prod_{i=1}^3 h_{p_i}^\beta (z)+\int_{-\infty}^{0} du_z \int_{u_z}^{2T-u_z} dv_z\ \bar h_{k}^0(z)\prod_{i=1}^3 h_{p_i}^0 (z)\bigg]\times\\
& \times\bigg[\frac{\alpha-1}{\alpha} \frac{2z^0}{i(k'+p_1+p_2+p_3)}+2\pi^2 \delta(k'+p_1+p_3+p_3)\delta(0)\bigg]\end{split}\end{equation}
The integral, which contains $z^0/(k'+p_1+p_2+p_3)$, can be calculated as follows:
\begin{equation}\begin{split} & \bigg\{\int_0^{T(1+\beta)} du \int_{\alpha u}^{2T-u} dv\ (u+v)\big[e^{ikv}-e^{ik\alpha u}\big]\prod_{i=1}^3\big[e^{-ip_i v}-e^{-ip_i\alpha u}\big]+\\
&+\int_{-\infty}^{0} du \int_{u}^{2T-u} dv\ (u+v)\big[e^{ikv}-e^{iku}\big]\prod_{i=1}^3\big[e^{-ip_i v}-e^{-ip_i u}\big]\bigg\}\cdot\frac{\alpha-1}{\alpha}\frac{1}{i(k'+p_1+p_2+p_3)}\sim\\
& \sim\bigg\{\int_{0}^\infty du\  e^{-i(-k+p_1+p_2+p_3)\alpha u}  \bigg[\frac{(2T-u)^2}{2}-\frac{(\alpha u)^2}{2}+(2T-u-\alpha u)\cdot u\bigg]+\\
& +\int_{-\infty}^0 du\  e^{-i(-k+p_1+p_2+p_3) u}  \bigg[\frac{(2T-u)^2}{2}-\frac{u^2}{2}+2T-2u\bigg] \bigg\}\cdot\frac{\alpha-1}{\alpha}\frac{1}{i(k'+p_1+p_2+p_3)}\sim\\
& = \frac{\alpha-1}{\alpha}\frac{T^2}{i(k'+p_1+p_2+p_3)}\bigg\{b_\alpha\cdot \operatorname{v.p.}\frac{1}{-k+p_1+p_2+p_3}+ c_\alpha\cdot\delta(-k+p_1+p_2+p_3)  \bigg\},\end{split}\end{equation}
where $b_\alpha$ and $c_\alpha$ are just constant coefficients, which depend on $\alpha$. The integral from eq. (\ref{eq:leading1}), which is proportional to $\delta(0)$, is equal to
$$2\pi^2\delta(0)\delta(k'+p_1+p_2+p_3)\bigg[\int_{0}^{T(1+\beta)} du_z \int_{\alpha u_z}^{2T-u_z} dv_z\ \bar h_{k}^\beta(z)\prod_{i=1}^3 h_{p_i}^\beta (z)+\int_{-\infty}^{0} du_z \int_{u_z}^{2T-u_z} dv_z\ \bar h_{k}^0(z)\prod_{i=1}^3 h_{p_i}^0 (z)\bigg].$$
The expression in square brackets has already been calculated in the section \ref{nkkbrokentraj}. So, using the result (\ref{eq:halfnkkbroken}), we obtain the answer:
$$-2\pi^2\delta(0)\delta(k'+p_1+p_2+p_3)\cdot \frac{2T(1-\alpha)}{\alpha}\bigg[ \operatorname{v.p.} \frac{i}{k+p_1+p_2+p_3}+\frac{1+2\alpha}{\alpha}\pi\delta(k+p_1+p_2+p_3)\bigg].$$
Therefore, the loop correction to $\kappa_{kk'}$ for the broken world-line is also proportional to $\lambda^2 T^2$:
\begin{equation}
\begin{split}
\label{eq:kappabroken}
\kappa_{kk'}\approx\   & \lambda^2 T^2\frac{1}{\sqrt{kk'}} \frac{\alpha-1}{\alpha} \int \bigg(\prod_{i=1}^3\frac{dp_i}{2\pi}\bigg) \frac{1}{p_1p_2p_3} \bigg\{ \frac{1}{i(k'+p_1+p_2+p_3)}\bigg[b_\alpha\cdot \operatorname{v.p.}\frac{1}{-k+p_1+p_2+p_3}+ \\
&+ c_\alpha\cdot\delta(-k+p_1+p_2+p_3) \bigg]-d_\alpha\cdot \delta(k'+p_1+p_2+p_3)\bigg[ \operatorname{v.p.} \frac{i}{k+p_1+p_2+p_3}+ \\
&+ \frac{1+2\alpha}{\alpha}\pi\delta(k+p_1+p_2+p_3)\bigg]\bigg\}  +\big\{k\leftrightarrow k'\big\},
\end{split}
\end{equation}
where $d_\alpha$ is a constant whose exact from is not very relevant for the estimates of the leading terms that we consider here. After getting rid of the last term, proportional to $\delta(k'+\sum p_i)$, one obtains that:
\begin{equation}\begin{split} \kappa_{kk'}\approx\  & \lambda^2 T^2\frac{1}{\sqrt{kk'}} \frac{\alpha-1}{\alpha} \int \bigg(\prod_{i=1}^3\frac{dp_i}{2\pi}\bigg) \frac{1}{p_1p_2p_3} \cdot \frac{1}{i(k'+p_1+p_2+p_3)}\bigg[b_\alpha\cdot \operatorname{v.p.}\frac{1}{-k+p_1+p_2+p_3}+\\
& +c_\alpha\cdot\delta(-k+p_1+p_2+p_3) \bigg]+\big\{k\leftrightarrow k'\big\}.\end{split}\end{equation}

\subsection{The mirror approaching the speed of light} \

In the case of the world-line (\ref{eq:exptraj}), the first integral over $u_w$ and $v_w$ in eq. (\ref{eq:generalf}) turns into
$$\int_{0}^{z^0-x(z^0)} du_w \int_{2t_{u_w}-u_w}^{2z^0-u_w} dv_w\ \bigg(e^{ik'v_w}-e^{ik'(2t_{u_w}-u_w)}\bigg)\prod_{i=1}^3 \bigg(e^{ip_i v_w}-e^{ip_i(2t_{u_w}-u_w)}\bigg)=$$
$$=\int_{0}^{z^0-x(z^0)}du_w\ \bigg[\frac{e^{i(k'+p_1+p_2+p_3)(2z^0-u_w)}-e^{i(k'+p_1+p_2+p_3)(2t_{u_w}-u_w)}}{i(k'+p_1+p_2+p_3)} +(2z^0-2t_{u_w})e^{i(k'+p_1+p_2+p_3)(2t_{u_w}-u_w)}+$$
\begin{equation}
\label{eq:zhiexp}
+\frac{e^{i(k'+p_1)(2z^0-u_w)}-e^{i(k'+p_1)(2t_{u_w}-u_w)}}{i(k'+p_1)}e^{i(p_2+p_3)(2t_{u_w}-u_w)}+...\bigg].\end{equation}
The ellipsis denote the remaining terms, obtained after the integration over $v_w$ of the following expressions (with $\sigma,\sigma_i=0,1$):

$$e^{i(\sigma k'+\sigma_1p_1+\sigma_2p_2+\sigma_3p_3)v_w}e^{i[ (1-\sigma) k'+(1-\sigma_1)p_1+(1-\sigma_2)p_2+(1-\sigma_3)p_3](2t_{u_w}-u_w)}.$$
The second term in (\ref{eq:zhiexp}) gives the leading contribution to $\kappa_{kk'}$ in the limit $T\to + \infty$:

\begin{equation}
\label{eq:someeq}
\begin{split}
& \int_{0}^{z^0-x(z^0)}du_w\ (2z^0-2t_{u_w})e^{i(k'+p_1+p_2+p_3)(2t_{u_w}-u_w)}= \\
& =\int_{0}^{z^0}dt_{u_w}\bigg[2-e^{-t_{u_w}/a} \bigg](2z^0-2t_{u_w})e^{i(k'+\sum_i p_i)a(1-e^{-t_{u_w}}/a)}.\end{split}\end{equation}
In different regions the integrand behaves as follows:
$$\operatorname{integrand}\approx \begin{cases}
\ \displaystyle 4(z^0-t_{u_w})e^{i(k'+\sum_i p_i)a}, & t_{u_w}\gg a,\\\\

\  \displaystyle  2\bigg(1+\frac{t_{u_w}}{a}\bigg)(z^0-t_{u_w})e^{i(k'+\sum_i p_i)t_{u_w}}, & t_{u_w}\ll a.
\end{cases}$$
Therefore, for those $z^0$, which are much greater than $a$, the integral (\ref{eq:someeq}) approximately equals $\displaystyle 2(z^0)^2e^{i(k'+\sum_i p_i)a}$. Thus, $\kappa_{kk'}$ is approximately equal to
$$\int_0^T dz^0 \int_{x(z^0)}^{z^0}dz\ \bar h_k^\beta(z^0,z) \prod_{i=1}^3 h_{p_i}^\beta(z^0,z)\cdot 2(z^0)^2\cdot e^{i(k'+\sum_i p_i)a}=$$
$$=2e^{i(k'+\sum_i p_i)a} \int_{0}^{T-x(T)} du \int_{2t_u-u}^{2T-u} dv\ (u+v)^2 \bigg(e^{ikv}-e^{ik(2t_{u}-u)}\bigg)\prod_{i=1}^3 \bigg(e^{-ip_i v}-e^{-ip_i(2t_{u}-u)}\bigg).$$
This integral is proportional to $T^4$ in the leading order. Indeed, let us show, how to calculate one of its parts:
\begin{equation}\begin{split} & \int_{0}^{T-x(T)} du\ u \int_{2t_u-u}^{2T-u} dv\  v\bigg(e^{ikv}-e^{ik(2t_{u}-u)}\bigg)\prod_{i=1}^3 \bigg(e^{-ip_i v}-e^{-ip_i(2t_{u}-u)}\bigg)\sim\\
& \sim \frac{1}{2}\int_{0}^{T-x(T)} du\ u \bigg[(2T-u)^2-(2t_u-u)^2\bigg] e^{-i(-k+\sum_i p_i)(2t_u-u)}=\\
& =\frac{1}{2}\int_{0}^{T} dt_u \bigg[2-e^{-t_{u}/a}\bigg]\bigg[2t_u-a(1-e^{-t_u/a})\bigg]\bigg[\big[2(T-t_u)+a(1-e^{-t_u/a})\big]^2-a^2(1-e^{-t_u/a})\bigg]\times\\
& \times e^{-i(-k+\sum_i p_i)a(1-e^{-t_u/a})}\sim \operatorname{const}\cdot e^{-i(-k+\sum_i p_i)a} \cdot T^4.\end{split}\end{equation}
So, the correction to $\kappa_{kk'}$ in the leading approximation as $T\to \infty$ is proportional to $\lambda^2 T^4$:
 \begin{equation}
 \label{eq:kappaexp}
 \kappa_{kk'}\propto \lambda^2 T^4\  \frac{e^{i(k+k')a}}{ \sqrt{kk'}}\cdot  \int \prod_{j=1}^3 \frac{dp_j}{2\pi} \frac{1}{p_1p_2p_3}.
 \end{equation}
 (see the discussion of the UV divergence, contained in this expression, in the section \ref{nkkexp}).

\section{One-loop corrections to the Keldysh propagator} \label{OneLoop}
\quad\quad Let us calculate the one-loop correction to the Keldysh propagator:
\begin{equation}
\label{eq:diagoneloop}
\Delta G^{K}_{xy}=\sum_{\sigma,\sigma_1=\pm1}
\begin{tikzpicture}

\begin{feynman}

 \vertex[label=${x, \sigma}$] (x);
    \vertex[label=${z, \sigma_1 }\ \ \ \ \ \ \ \ \ \ \ $] [right=2.5cm of x] (z);
    \vertex[label=${y, \sigma}$] [right=2.5cm of z] (y);

    \diagram* {

      (x) -- [plain] (z) -- [plain] (y)
      (z) -- [plain, out=130, in=50, loop, min distance=3cm] (z)

    };

   \end{feynman}
 \end{tikzpicture}
    \end{equation}
Using the expressions of the propagators through the mode functions (\ref{eq:throughmodes}), one can obtain the following results for $n_{kk'}$ and $\kappa_{kk'}$ for this loop in the limit $\displaystyle \frac{x^0+y^0}{2}= T  \to + \infty,\ \big|x^0-y^0\big|=\operatorname{const}$:
 $$n_{kk'}\propto \lambda \int d^2 z\ \bigg[\theta(x^0-z^0)\theta(z^0-y^0)-\theta(z^0-x^0)\theta(y^0-z^0)\bigg]\ h_k(z)\bar h_{k'}(z)  \int \frac{dp}{2\pi}\ h_p(z)\bar h_p(z)\bigg|_{x^0\approx y^0\approx T} \approx 0.$$
Thus, interestingly enough for any type of the mirror motion we obtain that the contribution to $n_{kk'}$ coming from one loop is always zero. However, that is not true for $\kappa_{kk'}$:

 \begin{equation}
 \label{eq:oneloopkappa}
 \kappa_{kk'}\propto \lambda \int d^2 z\  \theta(T-z^0)\  \bar h_k(z) \bar h_{k'} (z) \int \frac{dp}{2\pi}\ h_p (z)\bar h_{p} (z)=\lambda \int \frac{dp}{2\pi}\ \int d^2 z\  \theta(T-z^0)\  \bar h_k(z) \bar h_{k'} (z) h_p (z)\bar h_{p} (z).\end{equation}
Let us remind that the two-loop sunset diagram correction to $n_{kk'}$ looks as follows (\ref{eq:nkk}):
$$n_{kk'}\propto \lambda^2\int \bigg(\prod_{i=1}^3 \frac{dp_i}{2\pi} \bigg) \int d^2z\ \theta(T-z^0)\ {h}_k(z) \prod_{i=1}^3 h_{p_i}(z) \int d^2w\ \theta(T-w^0)\   \bar{h}_{k'}(w) \prod_{i=1}^3 \bar h_{p_i}(w),$$
and we have already calculated this expression in the section \ref{twoloopnkk}. Performing in the integral
$$\int d^2z\ \theta(T-z^0)\ {h}_k(z) \prod_{i=1}^3 h_{p_i}(z)$$
the following change of variables
$$k \to -k,\ p_1 \to -k',\ p_2 \to p,\ p_3 \to -p,$$
one can obtain the result for the one-loop correction (\ref{eq:oneloopkappa}):

\begin{equation}\kappa_{kk'}\propto \lambda \int \frac{dp}{2\pi} \frac{1}{\sqrt{2^4\cdot kk'p^2}}\cdot 2\pi\delta(k+k')\cdot\bigg(\sum_{\sigma,\sigma', \sigma_i=\pm 1} \sigma\sigma'\sigma_1\sigma_2\ \frac{1}{i(-\sigma k-\sigma' k'+\sigma_1 p-\sigma_2 p-i\epsilon)}\bigg) \approx 0 \end{equation}
for the mirror at rest, and

\begin{equation}
\begin{split}
\kappa_{kk'} \approx -\lambda\ \frac{2T(1-\alpha)}{\alpha} \int \frac{dp}{2\pi} \frac{1}{\sqrt{2^4\cdot kk'p^2}}\cdot \bigg[- \operatorname{v.p.} \frac{i}{k+k'}+\frac{1+2\alpha}{\alpha}\pi\delta(k+k')\bigg]+O(1)\end{split}
\end{equation}
for the broken world-line, and

\begin{equation}
\kappa_{kk'} \propto \lambda T^2 \int \frac{dp}{2\pi} \frac{1}{\sqrt{2^4\cdot kk'p^2}}\cdot e^{-i(-k-k'-i\epsilon)a}+O(T)\end{equation}\\
for the mirror, approaching the speed of light.

\section{Corrections to the four-point correlation functions} \label{FourPoint}\

In this section we also consider the following correction to the four-point correlation function in the limit when $x_0^1, x_0^2, x_0^3, x_0^4 \sim T \to +\infty$, while the distances $|x_0^i-x_0^j|$ for all $i< j$ are kept fixed:

\begin{equation}
\Delta G^{++--}_{x_1x_2x_3x_4}=\sum_{\sigma_i=\pm1}
\begin{gathered}
\begin{tikzpicture}
\begin{feynman}

 \vertex[label=${x_1, +}$] (x1);
    \vertex[label=0:$\ {z, \sigma_1 }$] [below right=1cm and 1.5cm of x1] (z);
    \vertex[label=${x_2, +}\ \ \ $] [below left=1cm and 1.5cm of z] (x2);
    \vertex[label=180:${w, \sigma_2}\ $] [right=4cm of z] (w) ;
    \vertex[label=${x_3, - }$] [above right=1cm and 1.5cm of w] (x3);
    \vertex[label=$\ \ \ \ \ {x_4, - }$] [below right=1cm and 1.5cm of w] (x4);

    \diagram* {

      (x1) -- [plain] (z) -- [plain] (x2)
      (z) -- [plain, out=55, in=125] (w)
      (z) -- [plain, out=-55, in=-125] (w)
      (x3) -- [plain] (w) -- [plain] (x4)

    };

   \end{feynman}
 \end{tikzpicture}
 \end{gathered}
\end{equation}
The sum of these diagrams is proportional to $$\lambda^2 \int d^2 z \int d^2 w\ \bigg\{ G^{+-}_{x_1 z} G^{+-}_{x_2 z} \big[G^{-+}_{zw}\big]^2 G^{+-}_{w x_3} G^{+-}_{w x_4}-G^{+-}_{x_1 z} G^{+-}_{x_2 z} \big[G^{--}_{zw}\big]^2 G^{--}_{w x_3} G^{--}_{w x_4}+$$
$$+G^{++}_{x_1 z} G^{++}_{x_2 z} \big[G^{+-}_{zw}\big]^2 G^{--}_{w x_3} G^{--}_{w x_4}-G^{++}_{x_1 z} G^{++}_{x_2 z} \big[G^{++}_{zw}\big]^2 G^{+-}_{w x_3} G^{+-}_{w x_4}\bigg\}.$$
By analogy with the formula (\ref{eq:occupanomal}), we represent this correction in the following form:
$$ \int \bigg(\prod_{i=1}^4 \frac{dk_i}{2\pi}\bigg)\ \bigg[n_{k_1k_2k_3k_4}^{(1)}\cdot\  h_{k_1}(x_1)h_{k_2}(x_2)\bar h_{k_3}(x_3)\bar h_{k_4}(x_4)+n_{k_1k_2k_3k_4}^{(2)}\cdot\  h_{k_1}(x_1)\bar h_{k_2}(x_2)\bar h_{k_3}(x_3)\bar h_{k_4}(x_4)+\ ...\bigg]$$
The last sum includes all possible terms with products of $h_{k_i}(x_i)$ and $\bar h_{k_j}(x_j)$. Let us calculate the function $n_{k_1k_2k_3k_4}^{(1)}$ in the approximation under consideration:
\begin{equation}
\label{eq:nkkkk}
\begin{split}
n_{k_1k_2k_3k_4}^{(1)}& \approx  \int \frac{dp}{2\pi} \int \frac{dp'}{2\pi}\int d^2z\ \bar h_{k_1}(z)\bar h_{k_2}(z) h_p(z)h_{p'}(z) \int d^2w\  \bar h_p(w)\bar h_{p'}(w)h_{k_3}(w) h_{k_4}(w)\times \\
& \times\bigg[\theta(z^0-T)\theta(w^0-T)-\theta(z^0-w^0)\theta(w^0-T)-\theta(w^0-z^0)\theta(z^0-T)\bigg]+\\
& +\int \frac{dp}{2\pi} \int \frac{dp'}{2\pi}\int d^2z\ \bar h_{k_1}(z)\bar h_{k_2}(z) \bar h_p(z) \bar h_{p'}(z) \int d^2w\ h_p(w)h_{p'}(w)h_{k_3}(w) h_{k_4}(w)\times\\
& \times\bigg[1-\theta(w^0-z^0)\theta(w^0-T)-\theta(z^0-w^0)\theta(z^0-T)\bigg]
\end{split}
\end{equation}
Simplifying the factor, containing theta-functions, from the first term above, we obtain:
\begin{equation}
\begin{split}
& \theta(w^0-T)\big[\theta(z^0-T)-\theta(z^0-w^0)\big]-\theta(w^0-z^0)\theta(z^0-T)=\\
&=\theta(w^0-T)\theta(z^0-T)\theta(w^0-z^0)-\theta(w^0-z^0)\theta(z^0-T)=\\
&=-\theta(T-w^0)\theta(w^0-z^0)\theta(z^0-T)=0.
\end{split}
\end{equation}
Consequently, only the second term in the formula (\ref{eq:nkkkk}) contributes to $n_{k_1k_2k_3k_4}^{(1)}$, which becomes equal after further simplification to:
\begin{equation}
\begin{split}
n_{k_1k_2k_3k_4}^{(1)}& \approx \int \frac{dp}{2\pi} \int \frac{dp'}{2\pi}\int d^2z\ \bar h_{k_1}(z)\bar h_{k_2}(z) \bar h_p(z) \bar h_{p'}(z) \int d^2w\ h_p(w)h_{p'}(w)h_{k_3}(w) h_{k_4}(w)\times\\
& \times\bigg[\theta(T-w^0)\theta(w^0-z^0)+\theta(T-z^0)\theta(z^0-w^0)\bigg]
\end{split}
\end{equation}

Notice that almost the same integrals have already been calculated in the section \ref{twoloopkappa} (see eqs. (\ref{eq:kappakk}), (\ref{eq:partkappa}), (\ref{eq:generalf})-(\ref{eq:kappabroken})), and they appeared to grow with time as some power of $T$. Similarly, every term in the last formula grows with time as $T^2$ in the case of the broken world-line with $\beta<1$ or as $T^4$ in the case of the mirror, approaching the speed of light. Thus, the correction to the four-point correlation function coming from $n_{k_1k_2k_3k_4}^{(1)}$ is proportional to $\lambda^2T^2$ or $\lambda^2T^4$, depending on the mirror world-line.

 \section{Conclusions and acknowledgements}\

 First of all, let us notice in conclusion that the exact Keldysh propagator in the interacting theory has the following form, when both points $x$ and $y$ have the same time coordinate $x^0=y^0$:
\begin{equation*} \begin{split}& G^{K}_{xy}\big|_{x^0=y^0}=G^{K (0)}_{xy}\big|_{x^0=y^0}+\\
& +\iint \frac{dk}{2\pi}\frac{dk'}{2\pi} \big[\braket{a_k^\dag a_{k'}} \bar{h}_k(x) h_{k'}(y)+\braket{a_k a_{k'}} h_k(x)h_{k'}(y)+\operatorname{c.c.}\big], \end{split}\end{equation*}
where $G^{K(0)}$ is a Keldysh propagator in the tree-level approximation, and operators $a_k^\dag,\ a_k$ are creation and annihilation operators in the interaction representation (consequently, they depend on time). Comparing this formula with the formula (\ref{eq:occupanomal}) for corrections to the Keldysh propagator, one notices that quantities $n_{kk'}$ are occupation numbers $n_{kk'}=\braket{a_k^\dag a_{k'}}$, while $\kappa_{kk'}$ are anomalous quantum averages $\kappa_{kk'}=\braket{a_k a_{k'}}.$ The fact that these quantities become non-zero indicates that the ground state of the system under consideration changes during its evolution.\\

To sum up, we have demonstrated in the article that in the two-dimensional massless scalar field theory with the null boundary condition on some time-like curve and with the $\lambda \phi^4$ self--interaction loop corrections to the Keldysh propagator (and, consequently, to the energy flux) grow with time, which can lead to a significant change in the value of the energy flux, found in the tree-level approximation. To make a clear statement about this effect, it is necessary to resum the leading corrections, coming from all loops, as it was done in already mentioned cases of de Sitter space scalar field theory and of scalar electrodynamics in the electric field background \cite{AkhmedovKEQ}-\cite{AkhmedovPopovSlepukhin}, \cite{Akhmedov:2014hfa}, \cite{Akhmedov:2014doa}.\\

I would like to acknowledge discussions with L.Astrakhantsev, A.Diatlyk and F.Bascone. I would like to thank E.Akhmedov for formulating these problems for me and shearing his insight into this subject. My work is done under the financial support of the state grant Goszadanie 3.9904.2017/BCh.

\end{document}